\documentclass[onecolumn]{emulateapj}
\usepackage{graphicx}
\usepackage[normalem]{ulem}
\newcommand{\src}{GS~1826$-$238}

\begin{document}

\title{Systematic Uncertainties in the Spectroscopic Measurements of
  Neutron-Star Masses and Radii from Thermonuclear X-ray
  Bursts. III. Absolute Flux Calibration}

\author{Tolga G\"uver$^{1,2}$, Feryal \"Ozel$^{3}$, Herman
  Marshall$^{4}$, Dimitrios Psaltis$^{3}$, Matteo Guainazzi$^{5}$, \& \\
  Maria D{\'{\i}}az-Trigo$^{6}$}

\affil{$^{1}$  Istanbul  University,  Science Faculty,  Department  of
  Astronomy and  Space Sciences,  Beyaz\i t, 34119,  Istanbul, Turkey}
\affil{$^{2}$ Istanbul University Observatory Research and Application
  Center, Beyaz\i t, 34119, Istanbul, Turkey}

\affil{$^{3}$  Department  of Astronomy,  University  of Arizona,  933
  N. Cherry Ave., Tucson, AZ 85721}

\affil{$^{4}$  Center for Space  Research, Massachusetts  Institute of
  Technology, Cambridge, MA 02139}

\affil{$^{5}$ European Space Astronomy Centre of ESA, PO Box 78,
Villanueva de la Ca\~nada, E-28691 Madrid, Spain}
 
\affil{$^{6}$ ESO, Karl-Schwarzschild-Strasse 2, D-85748 Garching bei
M\"unchen, Germany}
  
\begin{abstract}

Many techniques for measuring neutron star radii rely on absolute flux
measurements  in the  X-rays.  As  a  result, one  of the  fundamental
uncertainties  in these  spectroscopic  measurements  arises from  the
absolute  flux calibrations  of the  detectors being  used. Using  the
stable  X-ray  burster, \src,  and  its  simultaneous observations  by
Chandra HETG/ACIS-S and RXTE/PCA as  well as by XMM-Newton EPIC-pn and
RXTE/PCA,  we  quantify   the  degree  of  uncertainty   in  the  flux
calibration by  assessing the differences between  the measured fluxes
during  bursts. We  find that  the RXTE/PCA  and the  Chandra gratings
measurements agree with each  other within their formal uncertainties,
increasing our  confidence in  these flux measurements.   In contrast,
XMM-Newton  EPIC-pn   measures  14.0$\pm$0.3\%  less  flux   than  the
RXTE/PCA.  This is consistent with the previously reported discrepancy
with the flux measurements of EPIC-pn, compared to EPIC-MOS1, MOS2 and
ACIS-S detectors. We also  show that any intrinsic time dependent
  systematic  uncertainty that  may exist  in the  calibration of  the
  satellites  has already  been implicity  taken into  account in  the
  neutron star radius measurements.

\end{abstract}

\keywords{stars: neutron --- X-rays: bursts}

\section{Introduction}

Measurements of the masses and radii of neutron stars provide unique
information on the equation of state of matter at densities above the
nuclear saturation density. Extremely accurate mass measurements in
binary systems can be obtained through dynamical methods (see, e.g.,
Stairs 2006; Demorest et al.  2010).  However, without independent
determinations of the radii of these neutron stars, these measurements
provide only limits for the equation of state. In contrast, an
accurate measurement of the radius of even a single neutron star can
provide strong constrains on the equation of state of neutron star
matter (Lattimer \& Prakash 2001; \"Ozel \& Psaltis 2009; Read et
al. 2009).

Thermonuclear X-ray bursts observed from low mass X-ray binaries have
proven to be one of the most suitable events that can be used to
measure both the radii and the masses of neutron stars (see, e.g., van
Paradijs 1978, 1979; Damen et al. 1990; Lewin, van Paradijs, \& Taam
1993; \"Ozel 2006).  Using high signal-to-noise X-ray data and
time-resolved spectroscopy allows a measurements of the apparent
radius during the cooling segments of the bursts.  In addition,
time-resolved spectroscopy during a photospheric radius expansion
burst enables a measurement of the Eddington luminosity for each
source.  Combining these two measurements leads to an uncorrelated
determination of the neutron star mass and radius.

Thanks to the Rossi X-ray Timing Explorer (RXTE), which provided more
than 15 years of observations of bursts (Galloway et al.  2008), it
has now become possible to not only make neutron star mass and radius
measurements using observations of specific systems (see, e.g.,
\"Ozel, G\"uver, \& Psaltis 2009) but also to quantify the systematic
uncertainties in these measurements.  In Paper~I of these series
(G\"uver, Psaltis, \& \"Ozel 2012a), we analyzed more than 13,000
X-ray spectra from 12 different sources and showed that the errors in
the spectroscopic determination of neutron-star radii due to
systematic effects in the cooling tails of X-ray bursts are in the
range of $\simeq$3\%$-$8\%. In Paper~II (G\"uver, \"Ozel, \& Psaltis
2012b), we used all the archival photospheric radius expansion bursts
and found that for sources from which a large number of bursts are
observed, systematic uncertainties in the determination of the
Eddington flux are at the 5\%$-$10\% level.

In this paper, we focus on a third possible source of systematic
uncertainty, which is the absolute flux calibration of the
Proportional Counter Array (PCA) on board RXTE.  Absolute flux
uncertainties affect the normalization of the spectra obtained during
bursts and therefore the measurement of both the apparent radii and of
the Eddington flux.

Determining the absolute flux calibration of an X-ray detector
requires a major effort from any instrument team and all X-ray
observatories put large amount of resources into in-flight
calibrations of their detectors. However, these efforts are hindered
by the fact that almost all X-ray sources in the sky that may be used
as standard candles often show unpredicted variability over both long
and short time scales (see, e.g., Wilson-Hodge et al.  2011, and
references therein for the recent activity of the Crab Nebula).
Additionally, complications due to significantly different responses
of different detectors, both in terms of effective areas and the
energy ranges that the detectors are sensitive to, become a major
hurdle and restrict the inter-calibration efforts.  Furthermore, the
often unknown intrinsic spectral shape of the astronomical objects
being used for the calibration makes the problem even more complex.
Simultaneous observations of sources with well known intrinsic spectra
by two or more X-ray detectors that have different energy responses
allow us to better understand and quantify the systematic
uncertainties in the absolute flux calibrations of individual
detectors.

In the early days of high energy astrophysics, the Crab Nebula was
seen as the ideal stellar source for the calibration of various
instruments.  Compiling a number of rocket, balloon, and satellite
observations performed between 1964 and 1972, Toor \& Seward (1974)
concluded that the diffuse X-ray emission should be steady at least
for the time period they analyzed, since any variation in the pulsar
emission larger than 10\% at a few keV would have been detected. With
these limits, they assumed that the Crab is a steady source and
provided a ``universal'' power-law amplitude and photon-index with
their relevant uncertainties.  In the following years, these values
were often cited and used as a reference point to calibrate the
performance of various instruments.  Norgaard-Nielsen et al.\ (1994)
even proposed a specific mission called {\it eXCALIBur} to
continuously monitor the Crab Nebula using two pinhole cameras one
with a CCD and another with a proportional counter to measure the
spectrum of the nebula with an accuracy $<$2\% over the energy range
0.8$-$20~keV.  Such a mission could, in principle, be a reliable
inter-calibration reference for subsequent detectors.

Using the Crab as a standard source, Kirsch et al.\ (2005) compared
the XMM-Newton/EPIC-pn measurements of the spectrum of Crab with those
from all the other modern X-ray satellites. In particular, they found
that although the photon index inferred from XMM-Newton/EPIC agrees
within errors with the Toor \& Seward (1974) value, the normalization
derived from EPIC-pn is significantly lower than that from the rest of
the instruments.  They also noted that, with the new generation of
instruments, Crab is becoming a source that is too bright and too
extended to be used as a standard calibration source. In a similar,
later study, Weisskopf et al.\ (2010) investigated possible departures
from a single power-law of the intrinsic spectrum of Crab for
ROSAT/PSPC, RXTE/PCA, and the XMM-Newton/EPIC-pn, which would affect
the instrument calibrations. Using the high statistical quality of the
RXTE/PCA data, these authors concluded that they would have been able
to detect significant deviations from a simple power-law spectrum but
they did not find any evidence for such a deviation.

With the advancement of imaging detectors and considering the
brightness of the Crab for the onboard CCDs, searches for new
calibration sources and techniques started. In order to coordinate and
improve the efforts on inter-calibration, a group of instrument teams
formed The International Astronomical Consortium for High-Energy
Calibration (IACHEC, Sembay et al.  2010).  One of the recent results
of this group demonstrates that in the 2.0$-$7.0~keV range, EPIC-pn
measures systematically lower fluxes than the Chandra ACIS, by
11.0$\pm$0.5\% (Nevalainen et al.\ 2010), using a sample of 11 nearby
clusters of galaxies.  The difference is smaller yet still significant
when ACIS is compared with EPIC MOS1 and MOS2, with differences at
3.0$\pm$0.5\% and 6.0$\pm$0.5\%, respectively (Nevalainen et
al. 2010).

The spectroscopic measurements of neutron star masses and radii have
been predominantly performed with RXTE.  Launched in 1995 and operated
until 2012, RXTE/PCA has always been calibrated against the Crab
Nebula (Jahoda et al.\ 2006; Shaposnikov et al.\ 2012) together with
an onboard calibration source used mainly for channel-to-energy
conversion.  Throughout the mission, any possible variations in the
Crab, as the one reported by Wilson-Hodge et al.\ (2011), were assumed
to be averaged out (Shaposnikov et al.\ 2012); this issue will be
discussed further in Section~\ref{variable_source}.  Because the PCA
was intrinsically calibrated against the Crab and because it is a
non-imaging instrument, inter-calibration of the PCA has always been a
challenge, especially as more and more sources such as clusters of
Galaxies or similarly soft and relatively dim objects (i.e.,
RX~J1856.5$-$3754) started to be used as standard candles.

Time resolved X-ray spectra extracted during thermonuclear X-ray
bursts observed from low mass X-ray binaries offer a different
opportunity in determining the systematic uncertainties caused by
calibrations of different detectors.  Since the early studies of X-ray
bursts, it has been shown that the burst X-ray spectra can
statistically be modeled with blackbody functions (see, e.g., Swank et
al.\ 1977; Lewin, van Paradijs, \& Taam 1993; Galloway et al.\ 2008).
In Paper~I, we analyzed more than 13,000 X-ray spectra extracted from
the cooling tails of X-ray bursts observed from 12 different sources
and showed that a Planckian function can statistically fit the vast
majority of the observations. The high count rates of RXTE and the
large number of bursts allowed us to show that any deviation from a
Planckian spectrum is limited to $\leq 5\%$ in the 2.5$-$25.0~keV
range. This almost unique characteristic of burst spectra make them
ideal for testing our ability to convert countrate spectra to
bolometric fluxes.  Furthermore, because the obtained spectral
parameters are directly related to neutron star mass-radius
measurements, results obtained from an analysis of X-ray burst data
can be directly converted to uncertainties in the measurements of
neutron star masses and radii.

Using  simultaneous RXTE,  Chandra  X-ray  Observatory (Chandra),  and
XMM-{\it Newton} observations  of X-ray bursts from  \src, we quantify
the differences  in the measurements  of the apparent surface  area of
the thermal  emission when using different  detectors with independent
calibrations. We describe  our method in Section~2, which  is based on
predicting the observed Chandra  and XMM-{\it Newton} countrates using
time resolved  spectral parameters obtained from  RXTE/PCA.  Section~3
details  the analysis  of the  data from  different X-ray  satellites.
Finally, in Sections~4 and 5 we  present our results and discuss their
implications.

\section{Methodology}
\label{method}

The rapid  evolution of the  effective temperature during  the cooling
tails of thermonuclear  X-ray bursts demands very  short exposure time
observations for direct comparison of  spectral data that are obtained
simultaneously. The short exposures  require very high countrates and,
hence, detectors with  large photon collecting areas. This  has been a
unique  property of  RXTE/PCA  but  cannot be  achieved  by the  other
instruments  that  we  are  going  to  compare  its  results  against.
Therefore, we  formulated an approach  that does not rely  on directly
comparing  the  extracted  spectra   of  X-ray  bursts  simultaneously
observed  with  two (or  more)  X-ray  satellites. Instead,  we  first
analyze the RXTE data by  performing time resolved X-ray spectroscopy.
We extract spectra for short  intervals (0.25$-$2.0~s depending on the
countrate) and fit these data  with blackbody functions.  Based on the
obtained $\chi^2$ values, we then  calculate the likelihood for a wide
range of spectral  parameters that cover the  $\pm$5$-\sigma$ range of
the best fit values.  We use  this range of individual parameter pairs
(blackbody  temperature and  normalization) to  create simulations  of
Chandra or  XMM-{\it Newton} X-ray  spectra and then to  calculate the
total RXTE-predicted countrate  in a given energy  range. We calculate
the  uncertainty  in  these  countrates by  taking  into  account  the
posterior likelihoods of the  RXTE/PCA parameters for each simulation.
Finally,  we  compare  this  predicted countrate  with  the  countrate
actually observed  by Chandra or  XMM-{\it Newton} in the  same energy
range, taking into account all the associated uncertainties.

Assuming  that there  are  no calibration  uncertainties, the  Chandra
and/or XMM-{\it Newton}  count rates that were  predicted by analysing
the RXTE  observation should be  in perfect agreement with  the actual
count rates  observed with these  satellites.  A systematic  offset on
the other hand,  would imply a discrepancy in  the overall calibration
of either or both detectors.

It  is  evident  that  this  method allows  for  the  detection  of  a
systematic  uncertainty  caused by  the  overall  flux calibration  of
detectors and  specifically in the determination  of the normalization
in a particular  energy range.  However, the method does  not allow us
to constrain  any calibrational uncertainties in  the determination of
the spectral shape.

We searched  the archival  RXTE observations to  find the  best source
that  can be  used for  this  purpose.  An  ideal source  should be  a
frequent  burster  with  a  number of  simultaneous  observations  and
predictable  burst properties.   Two low  mass X-ray  binaries in  the
archive conform  to these conditions: the  so-called clockwork burster
\src\ and 4U~1636$-$536.   The problem with the latter  source is that
its bursts  do not last  long enough (only  about 20~s on  average) to
obtain sufficient amount of data to infer any systematic trends during
the cooling  tails of X-ray  bursts.  On  the other hand,  with bursts
that last  on average  130~s, \src\ provides  a significant  amount of
counts to be used for this purpose.

\subsection{\src}

\src\  was  first  discovered  with the  Ginga  satellite  (Makino  et
al.\   1988;   Tanaka\   1989).   First   conclusive   detections   of
thermonuclear  X-ray  bursts  from  this  source  came  from  BeppoSAX
observations  (Ubertini~et~al.~1997), although  X-ray bursts  may have
been observed from this source earlier by OSO~7 (Markert et al.\ 1977)
and  OSO~8 (Becker  et  al.\  1976).  Using  2.5  years of  monitoring
observations by BeppoSAX, during which  70 X-ray bursts were detected,
Ubertini~et~al.   (1999)  reported  that   the  X-ray  bursts  show  a
quasi-periodicity  of 5.76~hr,  a  remarkable  characteristic of  this
source. Later,  using 260  X-ray bursts observed  with the  Wide Field
Camera (WFC) on board BeppoSAX over 6 years, Cornelisse et al.\ (2003)
showed that the  burst recurrence time is indeed  stable but decreases
as the persistent flux increases.

A  total of  40  X-ray  bursts were  observed  with  the RXTE  between
1997-2003 (Galloway  et al.\  2008).  During  this time  interval, the
persistent  flux  steadily  increased  as the  burst  recurrence  time
decreased (Galloway et  al.\ 2008). Another remarkable  feature of the
X-ray bursts observed  from this source is  their similarity. Although
the  initial  decay  time-scale  and the  burst  fluence  show  slight
variations with persistent flux, the observed X-ray burst light curves
are nearly identical and reproducible (Galloway et al.\ 2004).

The nearly  identical light  curves of individual  bursts, predictable
recurrence   times,  and,   finally,  already   existing  simultaneous
observations  with  other  X-ray   satellites  (Chandra  and  XMM-{\it
  Newton}) make \src\  an ideal target to understand  and quantify the
systematic uncertainties in the measurement of neutron-star masses and
radii introduced by the uncertainties in the absolute flux calibration
of RXTE/PCA.

\section{Data Analysis}

In   this   study,   we   use  data   from   three   different   X-ray
satellites. \src\ was observed simultaneously  by Chandra and RXTE for
the first  time in 2002,  where three  X-ray bursts were  recorded. We
present  the  analysis  of  these  bursts  below.   Furthermore,  RXTE
observed  the system  in  2003 simultaneously  with XMM-{\it  Newton}.
During this observation,  7 X-ray bursts were detected  with RXTE.  We
analyze here  5 of these events  because, during one of  the remaining
bursts,  XMM-{\it  Newton}  was   affected  by  high  energy  particle
background  (burst \#28)  while  the other  burst  was only  partially
observed  and did  not have  high time  resolution data  with RXTE/PCA
(burst \#29). We present a log of these bursts in Table~\ref{bursts}.

\begin{deluxetable}{cccc}
\tablecolumns{4}   
\tablecaption{Analyzed X-ray Bursts} 
\tablewidth{0pt}
\tablehead{
\colhead{Burst ID\tablenotemark{a}}   &  \colhead{Start Time (MJD)\tablenotemark{a}} 
& \colhead{RXTE Observation ID} & \colhead{Simultaneous Obs.}}
  \startdata
22 & 52484.41775 & 70044-01-01-000 & Chandra \\
23 & 52484.56684 & 70044-01-01-00 & Chandra \\
24 & 52485.00672 & 70044-01-01-02 & Chandra \\
26 & 52736.53279 & 80048-01-01-00 & XMM-{\it Newton} \\
27 & 52736.66488 & 80048-01-01-010 & XMM-{\it Newton} \\
28\tablenotemark{b} & 52736.80048 & 80048-01-01-010 & XMM-{\it Newton} \\
29\tablenotemark{c} & 52738.34118 & 80048-01-01-04 & XMM-{\it Newton} \\
30 & 52738.47686 & 80048-01-01-04 & XMM-{\it Newton} \\
31 & 52738.61013 & 80048-01-01-07 & XMM-{\it Newton} \\
32 & 52738.74636 & 80048-01-01-07 & XMM-{\it Newton}
\enddata
\tablenotetext{a}{Burst  IDs and  burst start  times are  adopted from
  Galloway  et  al.   (2008).}   
\tablenotetext{b}{This burst is excluded from further analysis because,
  during the  burst XMM-{\it Newton} was exposed to  high background flaring
  activity.}
\tablenotetext{c}{This burst is excluded  from the analysis because it
  was only partly observed by the RXTE and high time resolution data
  modes of RXTE/PCA did not cover the burst (Galloway et al. 2008).\\}
\label{bursts}
\end{deluxetable}

\subsection{RXTE}

The RXTE observations  were analyzed following the  method detailed in
Paper~I.  Time  resolved, 2.5$-$25.0~keV X-ray spectra  were extracted
from science event mode data (125~$\mu$s time resolution and 64 energy
channels)  using the  ftool {\it  seextrct} from  all of  the RXTE/PCA
layers. For each  burst, spectra were integrated over 0.25,  0.5, 1 or
2~s time intervals so that the total number of counts in each spectrum
remain  roughly  the  same  while the  countrate  decreases  during  a
burst. A 16~s  X-ray spectrum was extracted from a  time interval just
prior to each  burst, as the spectrum of the  persistent emission, and
subtracted from the X-ray burst spectra as background. We took the PCA
deadtime correction into account  following the procedure described in
Paper I.

Response matrix files were created for each burst separately using the
PCARSP  version 11.7,  HEASOFT  6.7 and  HEASARC's remote  calibration
database. X-ray spectral analysis  was performed using the Interactive
Spectral  Interpretation System  (ISIS),  version  1.4.9-55 (Houck  \&
Denicola 2000)  and, for each  fit, a systematic uncertainty  of 0.5\%
was    included    as    suggested    by    the    RXTE    calibration
team\footnote[1]{http://www.universe.nasa.gov/xrays/programs/rxte/pca/
doc/rmf/pcarmf-11.7/}. We note that we also investigated the
possibility of photon pile-up in the RXTE/PCA data following Tomsick \&
Kaaret
(1998)\footnote[2]{http://heasarc.gsfc.nasa.gov/docs/xte/pca/tomsick\_98
. pdf} and Jahoda et al. (2006). During the decay of the X-ray bursts
investigated here, the observed count rates are $\approx$1600~cts/s/PCU,
which is far below the threshold of 10000 cts/s/PCU, where pile-up can
become important according to those studies. We, therefore, conclude
that photon pile-up in the PCA does not affect our results.

\begin{figure}
\centering
\mbox{
\includegraphics[scale=0.35]{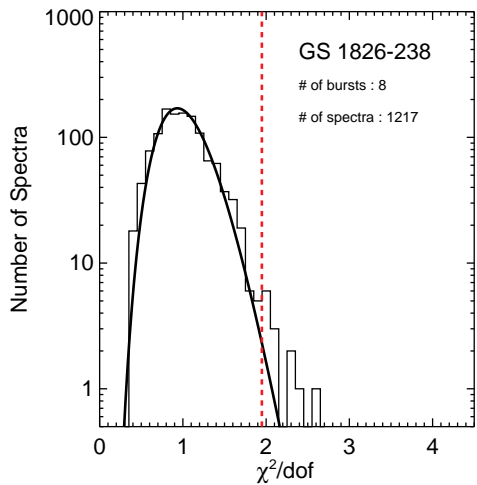}}
\caption{Histogram showing the $\chi^2$/dof distribution obtained from
  fitting 1217 X-ray spectra extracted from 8 thermonuclear X-ray
  bursts observed from \src.  The solid line shows the expected
  $\chi^2$/dof distribution for the number of degrees of freedom in
  the fits. The vertical dashed line correspond to the highest value
  of $\chi^2$/dof that we considered to be statistically acceptable. \\}
\label{chi_hist}
\end{figure}

\begin{figure*}
\centering
\includegraphics[scale=0.40]{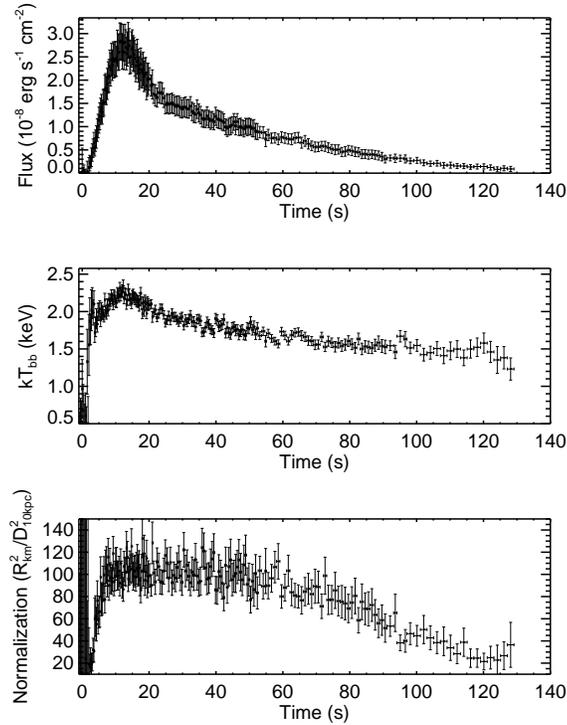}
\caption{Evolution of the spectral parameters during burst \#23
  as observed by RXTE. \\}
\label{rxte_burst23}
\end{figure*}

We  fit  each  spectrum  with  a blackbody  function  using  the  {\it
  bbodyrad}  model  (as defined  in  XSPEC,  Arnaud 1996).   Pinto  et
al.\ (2010) reported an analysis of high resolution X-ray spectroscopy
of  the  interstellar medium  toward  \src\  and found  an  equivalent
hydrogen  column density  of N$_{\rm  H} =  0.414 \times  10^{22}~{\rm
  cm}^{-2}$.   In  order  to  take  into account  the  effect  of  the
interstellar extinction in the X-ray  spectra, we used this value with
the {\it  phabs} model in XSPEC.   We set the elemental  abundances to
the    protosolar   abundances    of    Lodders   (2003),    following
Pinto~et~al. (2010).   The resulting histogram of  $\chi^2$/dof values
obtained from  the cooling tails  of all  the bursts and  the spectral
evolution during one  of the bursts (burst \#23) are  given in Figures
\ref{chi_hist}  and \ref{rxte_burst23}.   Similar  to  the results  we
obtained for a number of sources in Paper~I, the blackbody model could
adequately fit  the data with  a systematic uncertainty $\xi$  of only
$\approx~3\%$.

\subsection{Chandra X-ray Observatory}

Chandra  observed  \src\  in   2002  for  approximately  70~ks  (OBSID
2739). We analyzed  the data using the {\it  chandra\_repro} tool with
CIAO  version 4.3  and CALDB  4.4.5. The  {\it chandra\_repro}  script
automates the  data processing  following the standard  pipeline steps
but  applies  the  latest  available  calibrations.   Using  the  {\it
  dmextract}    tool     we    created    a     2.5$-$8.0~keV    range
background-subtracted source  lightcurve using only the  data from the
grating  arms  (+/- 1st  orders).   The  zeroth  order data  for  this
observation is saturated  due to the high countrate of  the source and
the nominal frame time of 1.7~s.

\begin{figure*} \centering
\includegraphics[scale=0.40]{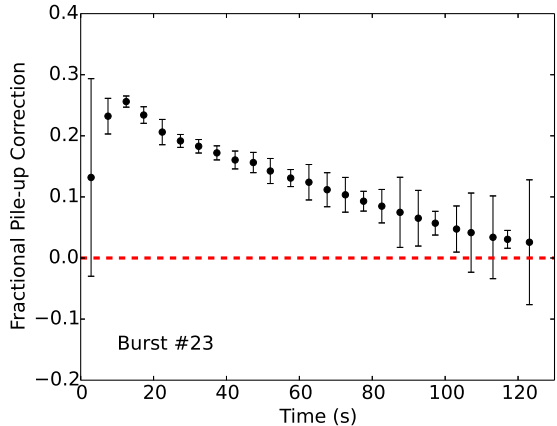} 
\caption{Evolution of the pile-up correction applied to the
predicted Chandra countrates based on the RXTE spectra as 
burst \#23 proceeds. For clarity the data are binned to 5~s 
intervals.}
\label{ch_pileup} \end{figure*}

Even for the grating arms, pile-up  was significant during the peak of
the bursts. In order to correctly carry out the comparison between the
RXTE and Chandra observations, we opt  to correct for pile-up the {\it
predicted} Chandra countrates based on  the RXTE spectra, as opposed to
correcting   the  observed   Chandra  countrates.    Because  RXTE
observations  have  better statistics,  this  forward  folding of  the
correction  does  not  unnecessarily inflate  the  uncertainties. 
We applied  a  pile-up correction  procedure  that  is described  in the
Appendix,  which  was developed  based  on  the understanding  of the
detector  properties and  calibrated  against  independent data. The
evolution of the applied correction is shown in Figure~\ref{ch_pileup}
for burst \#23.  At the burst peak, photon pile-up causes a net loss of
25\% of the flux, which can be recovered by the pile-up correction
model that was developed independently of the present data (see the
Appendix). Nevertheless, the  rise and the peak of  each burst, which
approximately correspond to the first  30~s, are most heavily affected
by pile-up. Furthermore, the rise and  the peak are also  the periods
where the intrinsic spectrum may  show some deviation from a Planckian
shape; this  is  expected  from theoretical  model spectra and  was
discussed in  Paper  I  using RXTE  observations. Because  of these
reasons, we exclude the  first 30~s  of all  Chandra bursts from our
analysis such that in the time interval that we use, the pile-up 
fraction decreases to 2\%.

We  used  the  {\it  mkgrmf}   and  {\it  fullgarf}  tools  to  create
observation  specific response  and auxiliary  response files  for the
first orders of  the high energy transmission grating  (HETG) and used
these when creating  the simulations based on  the spectral parameters
determined from time resolved X-ray spectroscopy.

\subsection{XMM-{\it Newton}}

XMM-{\it Newton} observed \src\ in  2003 for approximately 200~ks with
two different back-to-back exposures  that lasted 107,850 and 91,916~s
(OBSIDs : 0150390101, 0150390301).  The observation was performed with
EPIC-pn in TIMING mode and the  EPIC-MOS detectors were turned off. We
analyzed   the   EPIC-pn   data   using   the   SAS   version   11.0.0
(xmmsas\_20110223\_1801-11.0.0) and  the calibration files as  of June
2011.  We calibrated the observation  with the {\it epproc} tool.  Due
to the  high energy particle  background we  could only use  the first
67~ks  of the  first  observation and  $\approx$80~ks  of the  second.
Similar to the Chandra Observation, we extracted background subtracted
source light curves  in the 2.5~$-$~10.0~keV range using  the SAS tool
{\it             epiclccorr}.              The             observatory
guide\footnote[2]{http://xmm.esac.esa.int/external/xmm\_user\_support/documentation/uhb/epicmode.html}
gives a  limit of 450~ct/s for  telemetry of the EPIC-pn  timing mode,
which can be increased if some of the other instruments are turned off
as in  this case.  In  the burst  lightcurves that were  observed with
XMM-{\it Newton}, two  gaps that last approximately  10~s (see Figures
\ref{comp_xmm1} and  \ref{comp_xmm2}) are seen. These  are most likely
due  to this  limit.   While creating  the  simulations we  calculated
observation specific  effective area  and redistribution  matrix files
using the SAS tools {\it rmfgen} and {\it arfgen}.

\begin{figure*} \centering
\includegraphics[scale=0.40,angle=270]{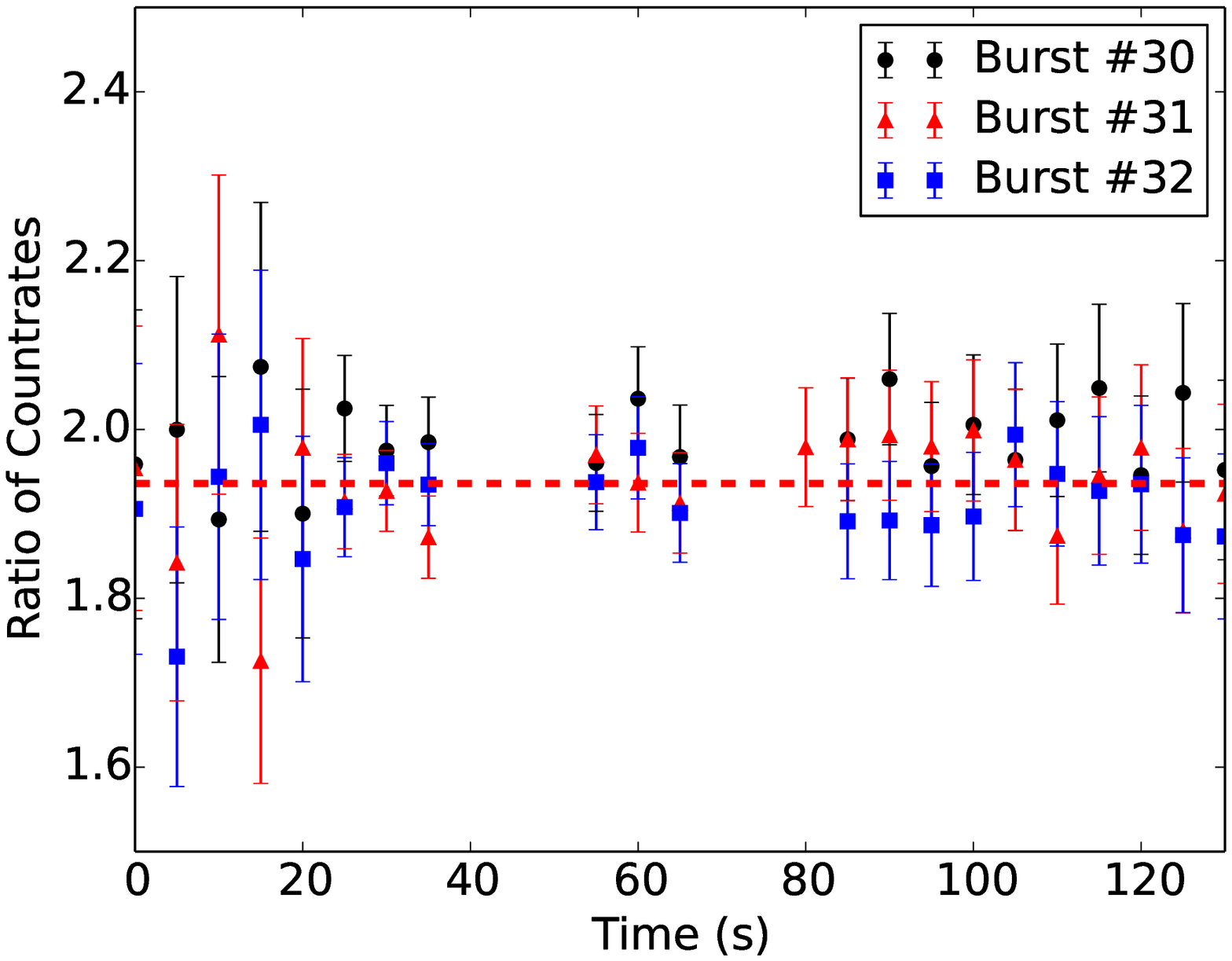}
\caption{ Ratio of countrates obtained when the full PSF is used to the
countrates when the central two pixels are excised, for three bursts
observed with XMM-Newton. For clarity the lightcurve is binned to 5~s
intervals} \label{xmm_pileup} \end{figure*}

In order to explore the possibility of pile-up in the EPIC-pn data, we
created two event lists: one including the full PSF for the source and
another  where  the central  two  pixels,  which  should be  the  most
affected from pile-up,  are excised. If pile-up  were significant, the
ratio of  the countrates in  the two event  lists would evolve  as the
count rate changes during the burst  because the two event lists would
be subject  to different  amounts of  pile-up.  The  ratio of  the two
lightcurves is  shown in Figure~\ref{xmm_pileup} for  bursts \#30, 31,
and 32. As  expected, a large fraction of counts  are collected in the
central  two pixels,  making this  ratio approximately  equal to  two.
However, as the burst eveolves, the  ratio remains constant at the few
percent level, implying that the pile-up effect will be at most of the
same order.  We consider this as  strong evidence for the  lack of any
pile-up effect on the measured fluxes. Nevertheless, in order to avoid
any residual  effects of  photon pile-up  and any  possible deviations
from a  Planckian shape, as  with the  Chandra bursts, we  exclude the
first  30~s of  each  burst when  determining  the systematic  offsets
between the  instruments.  We show  these time intervals in  the plots
for comparison and completeness.

\section{Comparison Between RXTE, Chandra, and XMM-{\it Newton}}
\label{res}

\begin{figure*} \centering \includegraphics[scale=0.40]{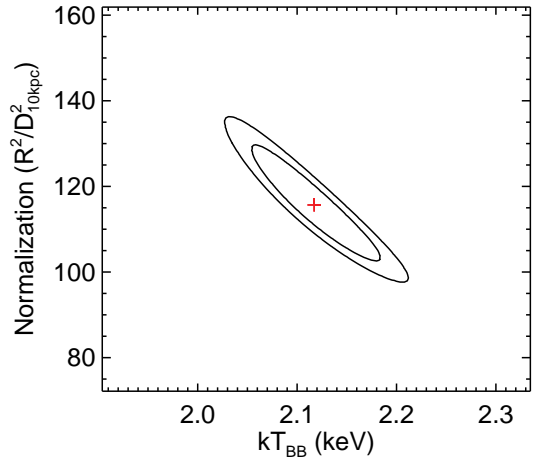}
\caption{Two dimensional 68\% and 95\% confidence contours for the
  spectral parameters obtained from the fit to an X-ray spectrum
  extracted approximately 9.7~s after the start of burst \#26.  In the
  calculation of the uncertainty in the total predicted countrate, the
  whole range of parameter values are taken into account together with
  their posterior likelihood.}
\label{example_conf} 
\end{figure*}

We followed the method  outlined in Section~\ref{method} for comparing
all of the X-ray bursts  observed simultaneously with RXTE and Chandra
or with RXTE and XMM-{\it Newton}.  Briefly, our analysis steps are as
follows:

\begin{enumerate} 

\item{We first obtained the best fit values of the parameters for each
  burst spectrum observed with RXTE/PCA.  We then calculated the
  two-dimensional confidence contours for the resulting color
  temperature and the blackbody normalization parameters covering a
  $\pm5-\sigma$ range from the best fit values.  An example confidence
  contour is shown in Figure~\ref{example_conf} for the spectrum
  obtained approximately 9.7~s after Burst~\#26 started. Note
    that we show 68\% and 95\% confidence contours in the figure but
    use the entire likelihood in the calculations.}

\item{For  each  pair  of   spectral  parameters,  we  calculated  the
  predicted countrates  for the  Chandra gratings (only  the $+/-$~1st
  orders for HEG and MEG) in  the 2.5$-$8.0~keV range and for XMM-{\it
    Newton}/EPIC-pn countrates in the  2.5$-$10.0~keV range (i.e., the
  energy range that matches the RXTE/PCA sensitivity). For the case of
  the  Chandra  simulations, we  applied  the  pile-up correction,  as
  described in the previous section.}

\item{In order to take into  account the uncertainties in the RXTE/PCA
  modeling, we used the posterior  likelihoods over the parameters for
  each  spectral  fit  to  calculate  the  uncertainty  in  the  total
  predicted countrates.}
  
\item{Finally,  we  compared  the   observed  Chandra  HETG/ACIS-S  or
XMM-{\it  Newton}/EPIC-pn  countrates  to the  countrates  predicted
using the  RXTE/PCA spectral fits.   For this comparison,  we binned the
countrates to match the time resolution of each data set and
subtracted the persistent flux level of the source from the burst
emission. During the Chandra observation, the nominal  frame time was
1.7~s. The time resolution  of  the  EPIC-pn   observation  was  much 
higher.   We, therefore, binned  the EPIC-pn  data to  match the 
varying exposure time of RXTE spectra.}

\end{enumerate}

\begin{figure*} \centering
\includegraphics[angle=270,scale=0.40]{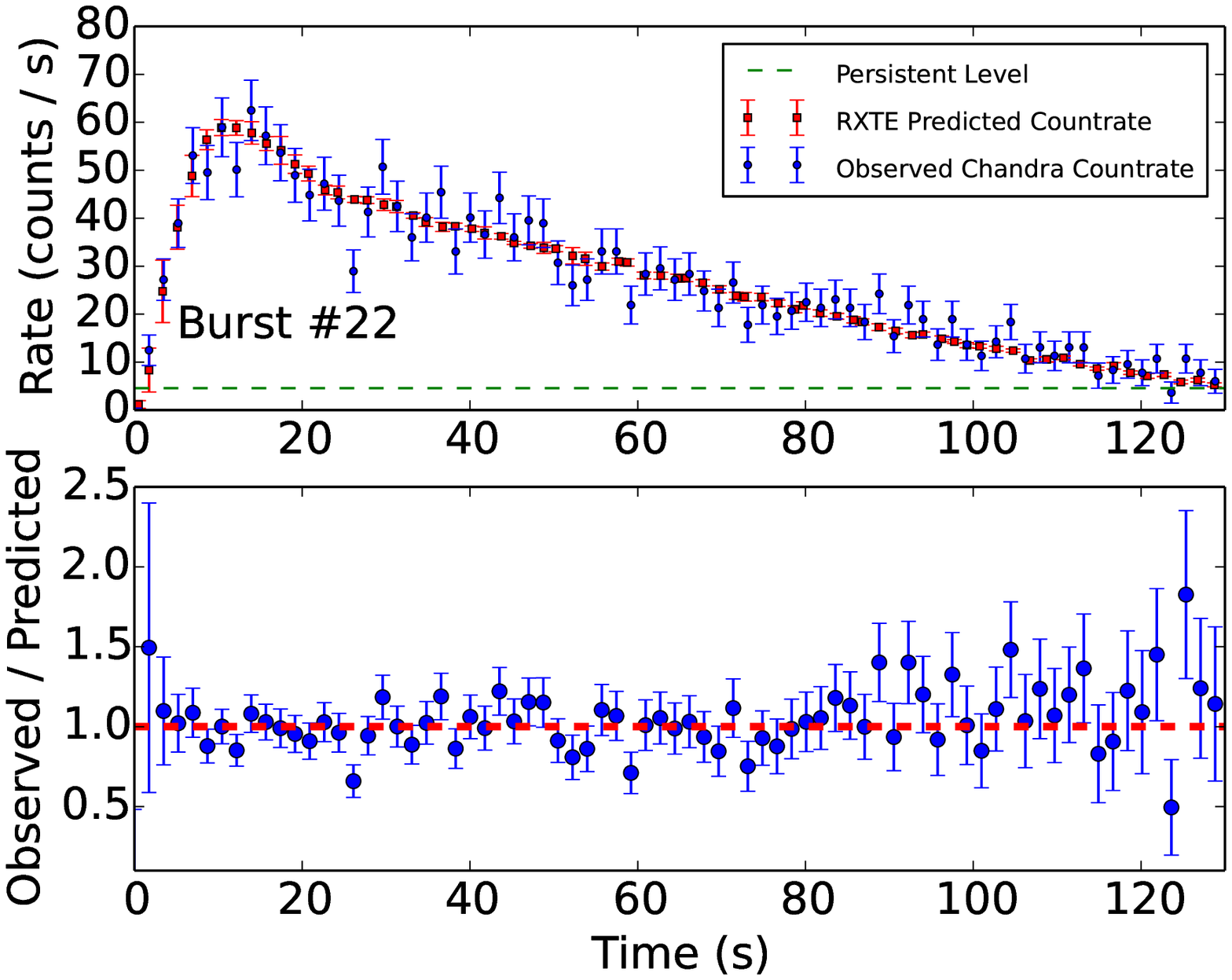}
\includegraphics[angle=270,scale=0.40]{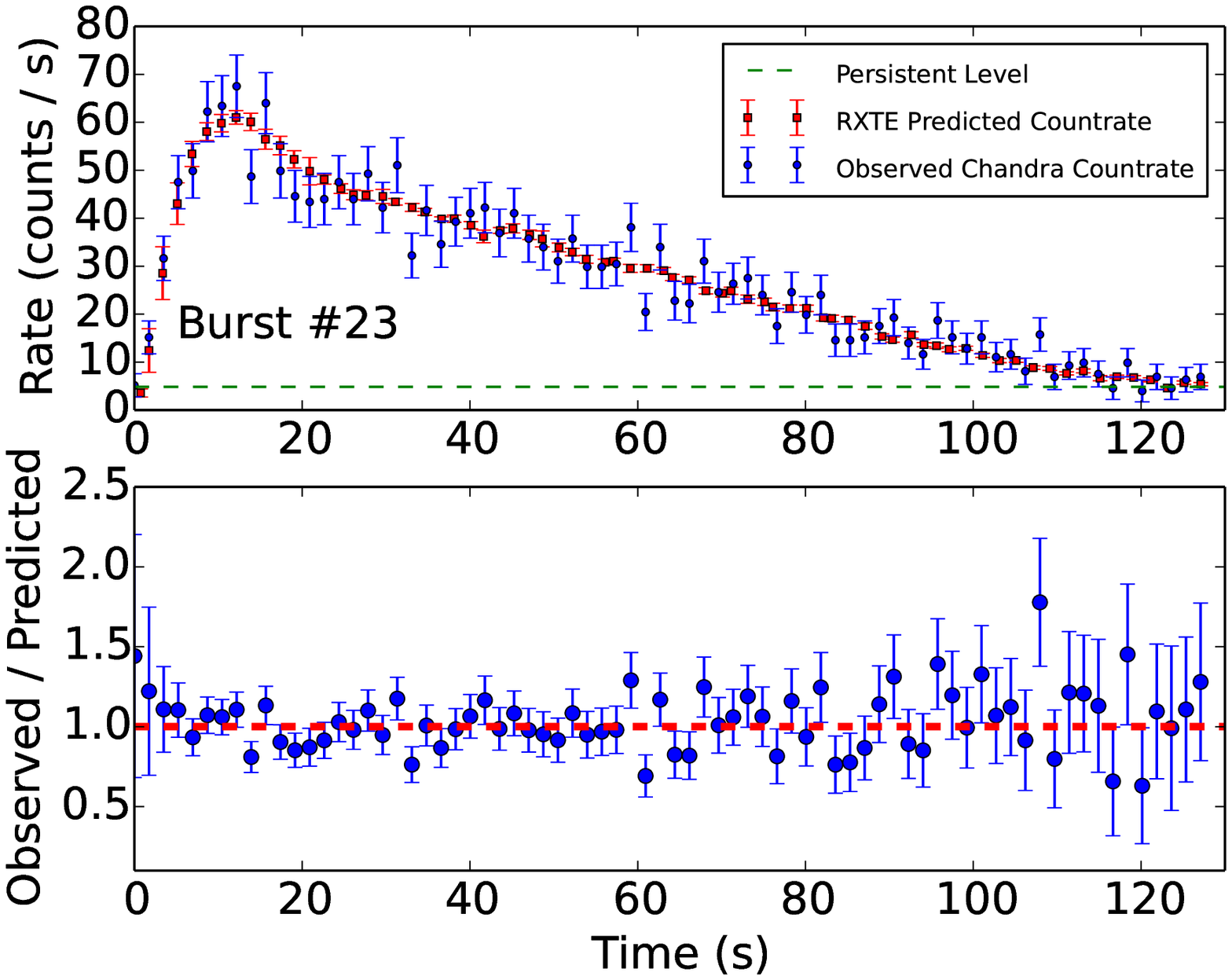}
\includegraphics[angle=270,scale=0.40]{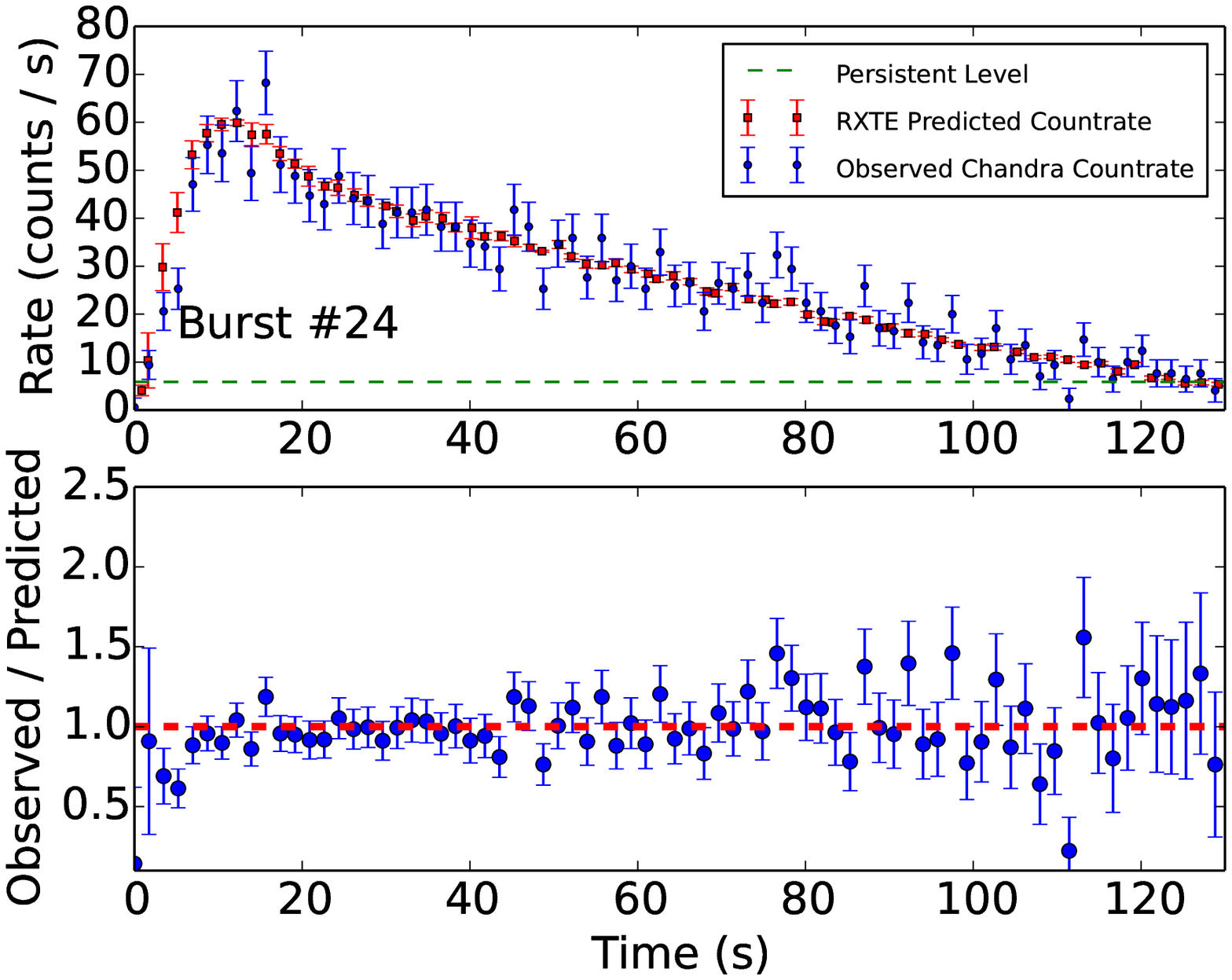} \caption{Comparison of
the countrates during individual  X-ray bursts as observed by Chandra
HETG/ACIS and as predicted using the spectral parameters inferred during
the simultaneous  RXTE/PCA observations. Note that the persistent
flux level of \src ~is shown with green dashed lines and it is
subtracted from the observed burst lightcurve. Lower  panels  show  the
 ratios   of the  observed and  predicted countrates.  The  error  bars 
in   the lower  panels reflect  the uncertainties of both  the Chandra
observations and  of the modeling of the RXTE observations.}
\label{comp_ch} \end{figure*}

\begin{figure*}
\centering
\includegraphics[angle=270,scale=0.40]{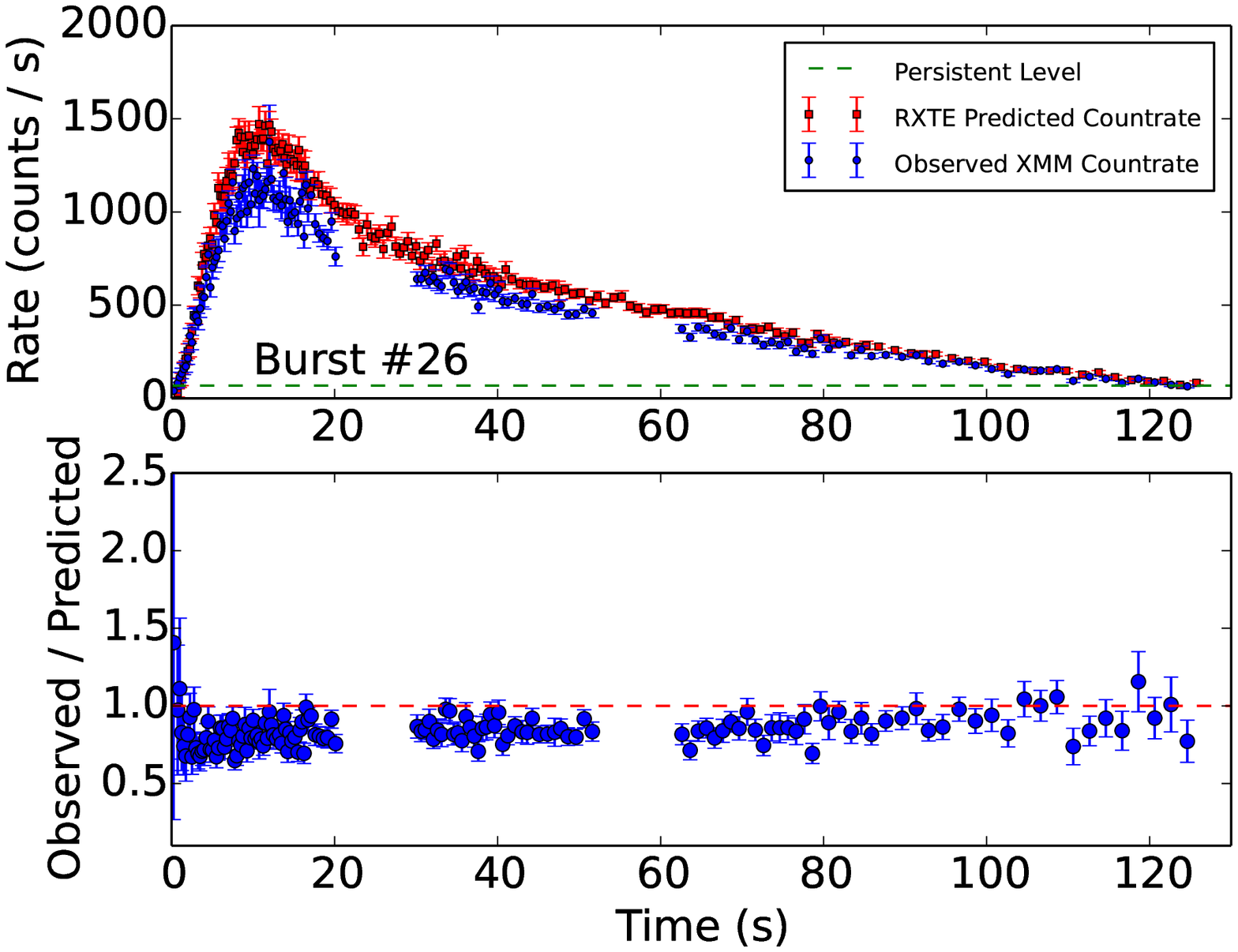}
\includegraphics[angle=270,scale=0.40]{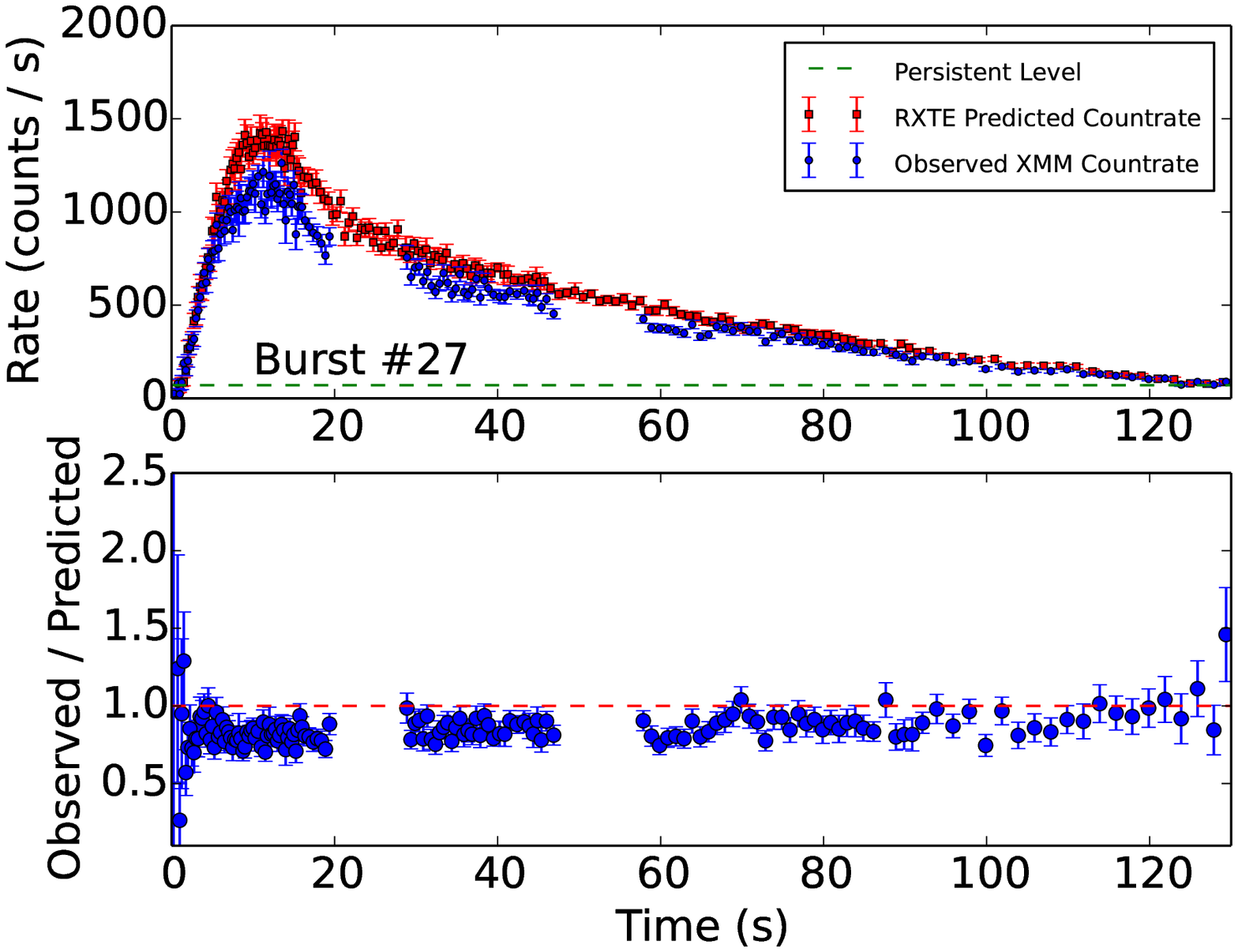}
\includegraphics[angle=270,scale=0.40]{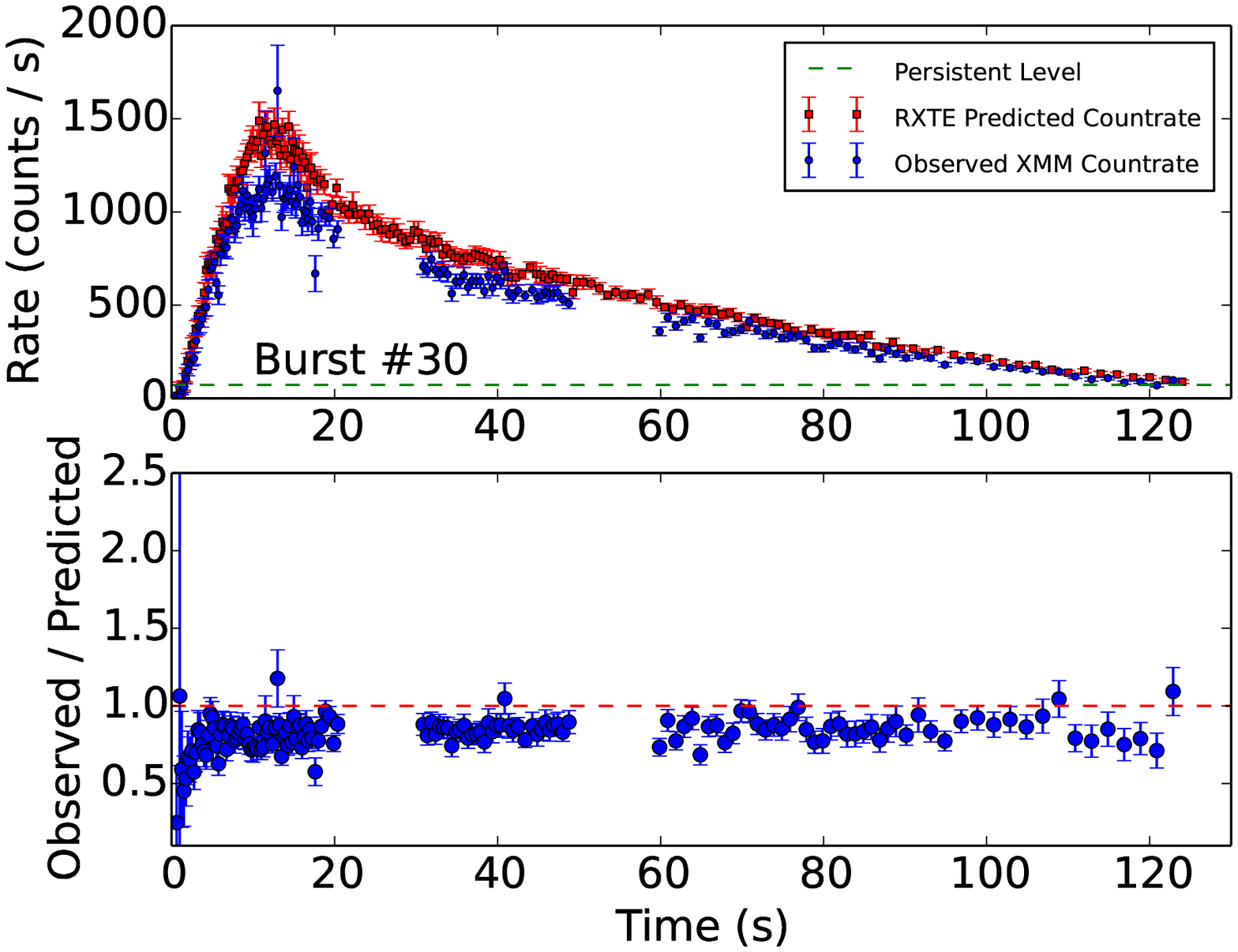}
\caption{Same as Figure~\ref{comp_ch} but for the X-ray bursts \#26,
  \#27, and \#30 observed simultaneously with RXTE/PCA and XMM-{\it Newton}
  EPIC-pn.}
\label{comp_xmm1}
\end{figure*}

\begin{figure*}
\centering
\includegraphics[angle=270,scale=0.40]{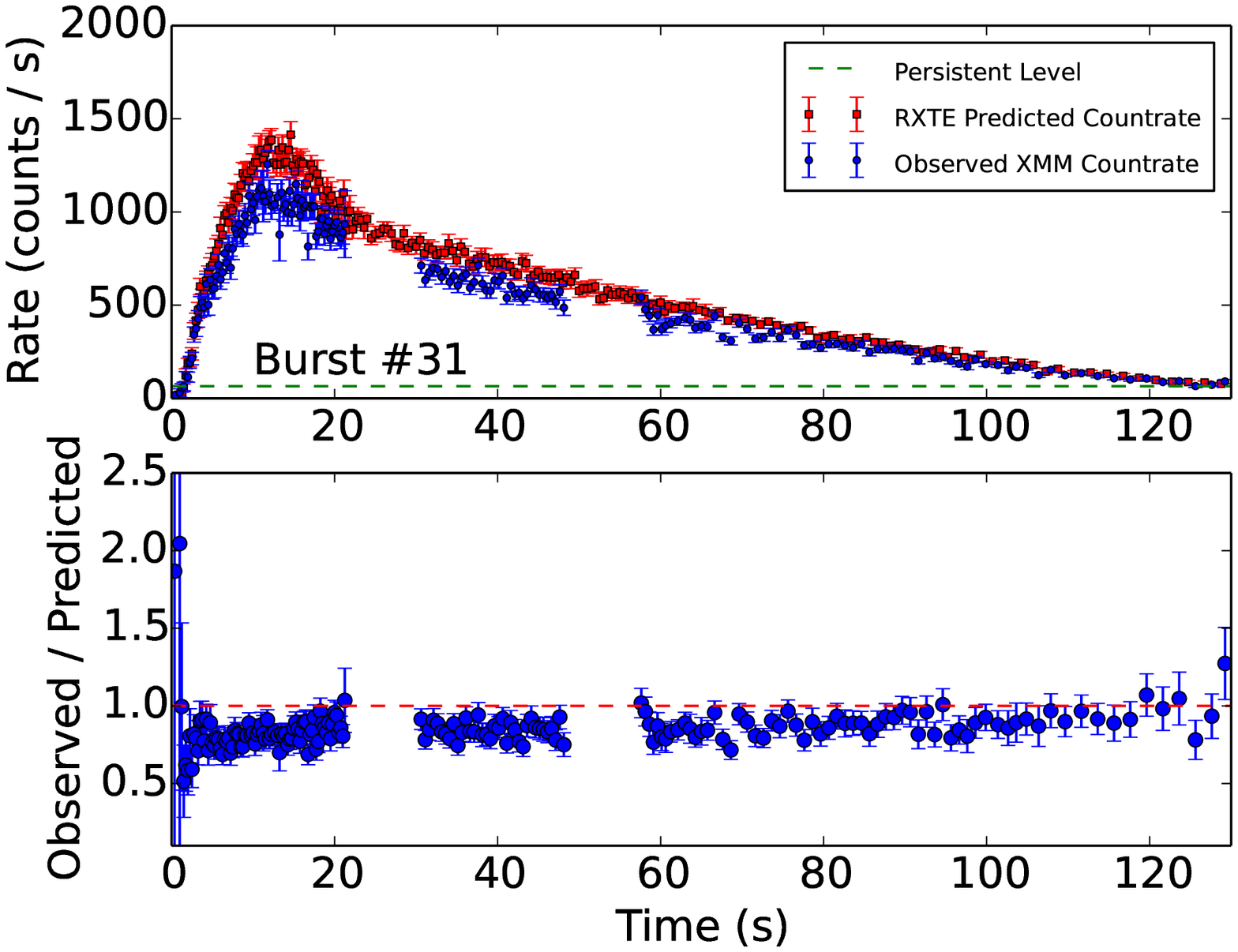}
\includegraphics[angle=270,scale=0.40]{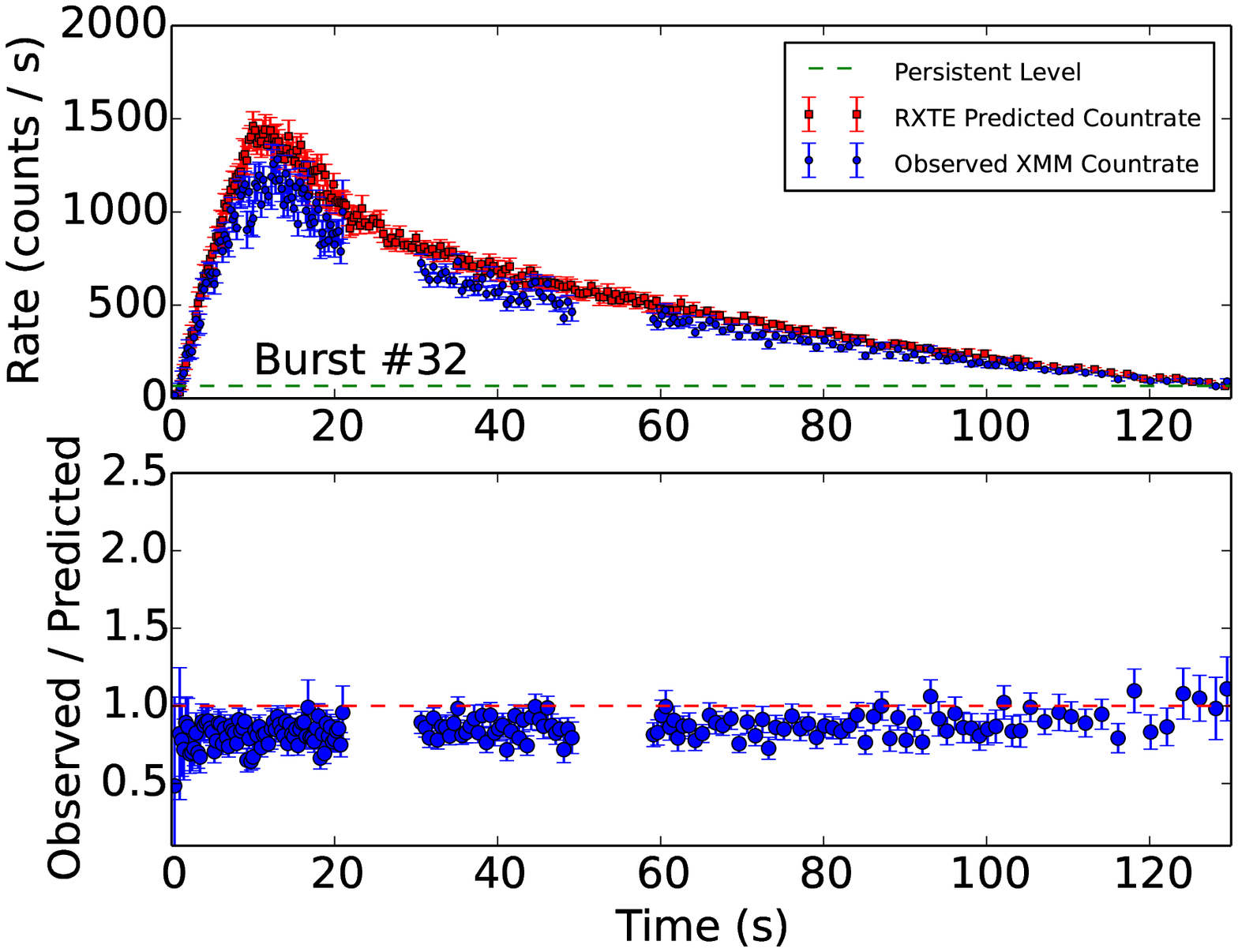}
\caption{Same as Figure~\ref{comp_ch} but for the X-ray bursts \#31
  and \#32 observed simultaneously with RXTE/PCA and XMM-{\it Newton} EPIC-pn.}
\label{comp_xmm2}
\end{figure*}

In  Figures~\ref{comp_ch},  \ref{comp_xmm1}, and  \ref{comp_xmm2},  we
show  the time  evolution of  the source  countrate during  each X-ray
burst   simultaneously  observed   with  RXTE   and  Chandra   in  the
2.5$-$8.0~keV  range  or  with  RXTE   and  XMM-{\it  Newton}  in  the
2.5$-$10.0~keV  range.  In principle,  all  of  the calculated  ratios
should be equal  to unity within their formal  uncertainties, if there
are no systematic calibration differences.  It is already evident from
this figure that this is indeed  the case between RXTE and Chandra but
that  there  is  a  systematic calibration  offset  between  RXTE  and
XMM-{\it Newton}.

\begin{figure*}
\centering
\includegraphics[angle=270,scale=0.40]{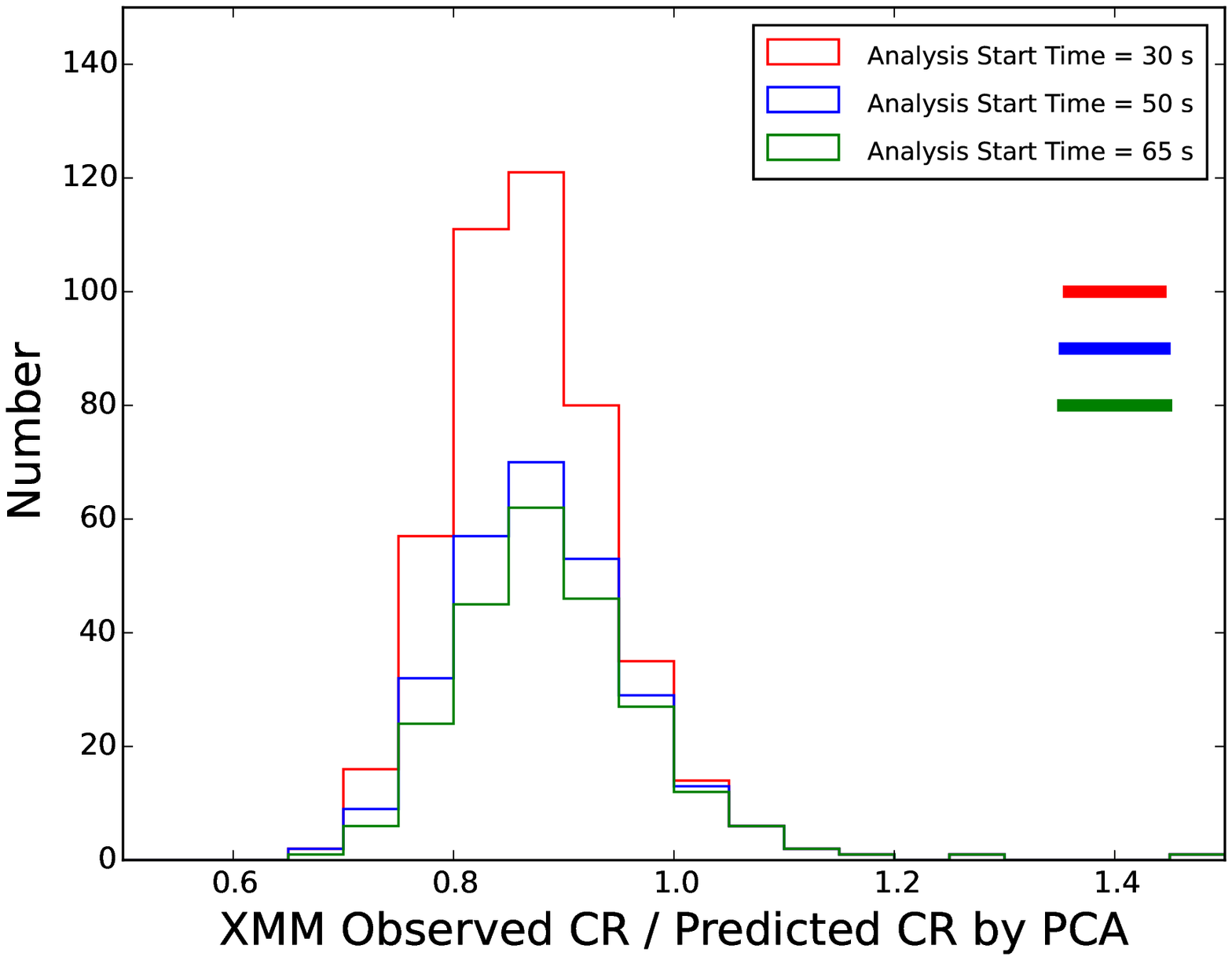}
\caption{Distributions of the ratios of the predicted and observed
  countrates for XMM-{\it Newton}/EPIC-pn, as measured from all of the
  five bursts. The three histograms correspond to measurements
  starting at 30, 50 and 65~s after the start of each burst. The three
  error bars represent the average statistical errors in the
  measurements for each segment.}
\label{phtest2}
\end{figure*}

To  explore in  more detail  the calibration  differences between  the
three instruments and whether  such differences evolve with countrate,
we  performed two  analyses. First,  for the  XMM-{\it Newton}/EPIC-pn
data, for which we infer the largest systematic calibration offset, we
created  histograms  of  the  ratios of  the  predicted  and  observed
countrates with varying start times  compared to the burst start times
(see  Figure \ref{phtest2}).  If  there were  a significant  countrate
dependence of the calibration offset, we would have seen a significant
change in  the distribution  of calculated ratios  at later  stages of
these bursts.  However, it  is evident from  Figure~\ref{phtest2} that
there is no such time dependent change in the ratios.

Second, in  order to quantify  the level  of offset between  the three
instruments, we followed the Bayesian  procedure described in Papers I
and II and outlined below for completeness.

We calculated the ratio $f_i$ between the observed and predicted
countrates for each time bin in the 30$-$120~s interval after the
start of every burst together with its formal uncertainty.  We assume
that the uncertainty in each measurement is described by a Gaussian
with a centroid $f_{0,i}$ and a dispersion $\sigma_{i}$ i.e.,
\begin{equation}
P_i(f|f_{0,i},\sigma_i)=C_i\exp\left[-\frac{(f-f_{0,i})^2}{2\sigma_i^2}\right]\;,
\end{equation}
where $C_i$ is a set of normalization constants.

We will also  assume that the ensemble of these  ratios for each burst
is drawn from  a Gaussian distribution with a centroid  at $f_0$ and a
dispersion $\sigma$, i.e.,
\begin{equation}
  P(f;f_0,\sigma)=C\exp\left[-\frac{(f-f_0)^2}{2\sigma^2}\right]\;,
\end{equation}
where $C$ is a proper normalization constant. This distribution is our
model for the underlying systematic flux calibration offset between
the instruments. If there are no calibration differences, the centroid
should be equal to one, to within $\sigma_i/\sqrt{N}$, where $N$ is
the number of measurements, and the dispersion should tend to zero.  A
significant offset in the flux calibration between two instruments
will move $f$ away from unity, but keep the dispersion consistent with
zero. On the other hand, a large dispersion cannot be produced by an
overall flux calibration offset. Instead, it likely points to effects
such as an inadequate modeling of the pile-up correction, of the
energy response of the detector(s), or of the astrophysical properties
of the source. Our goal is to calculate the posterior likelihood for
the parameters $f_0$ and $\sigma$, $P(f_0,\sigma|$data$)$, given an
ensemble of $N$ measurements.

\begin{figure*}
\centering
\includegraphics[scale=0.35]{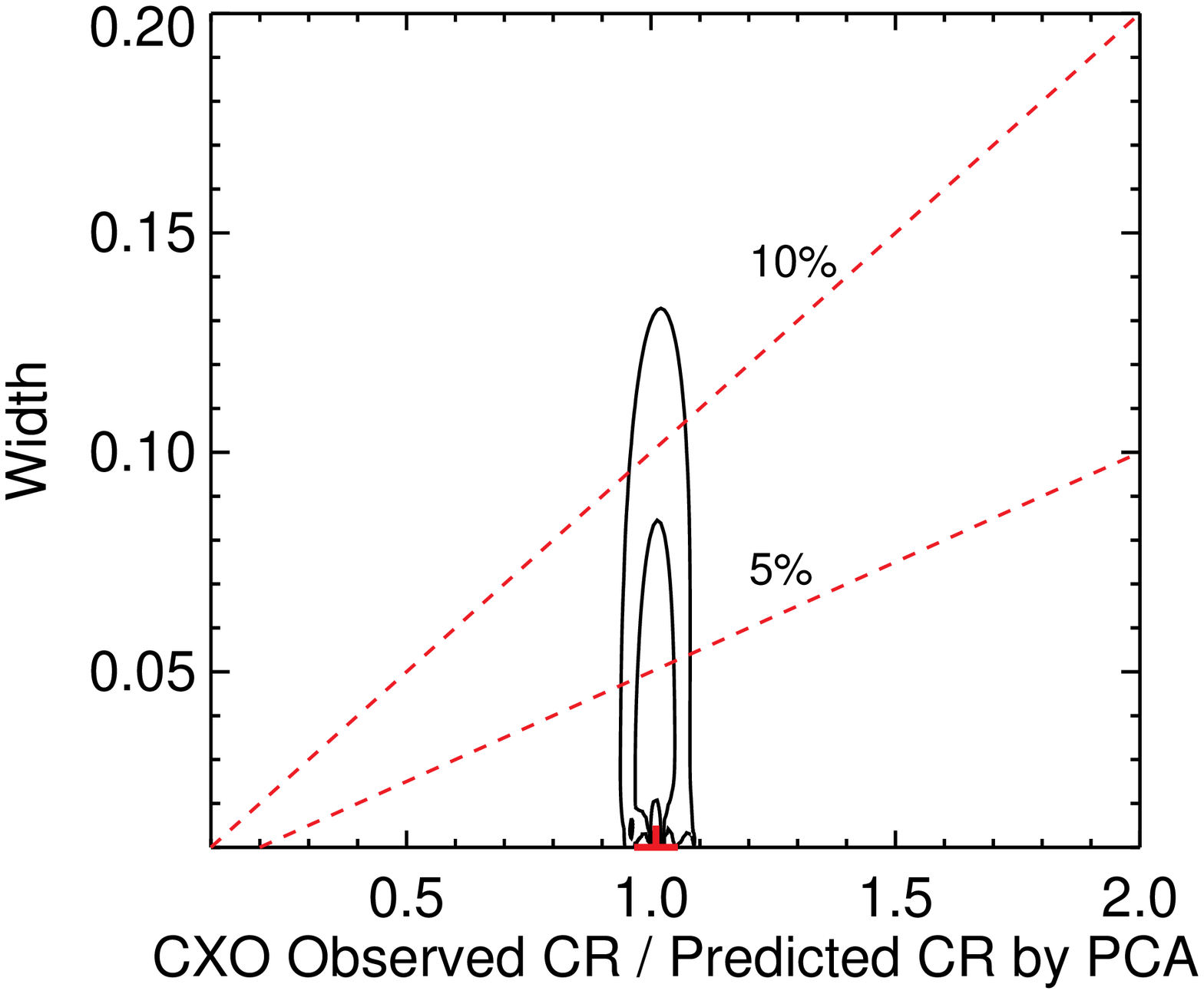}
\includegraphics[scale=0.35]{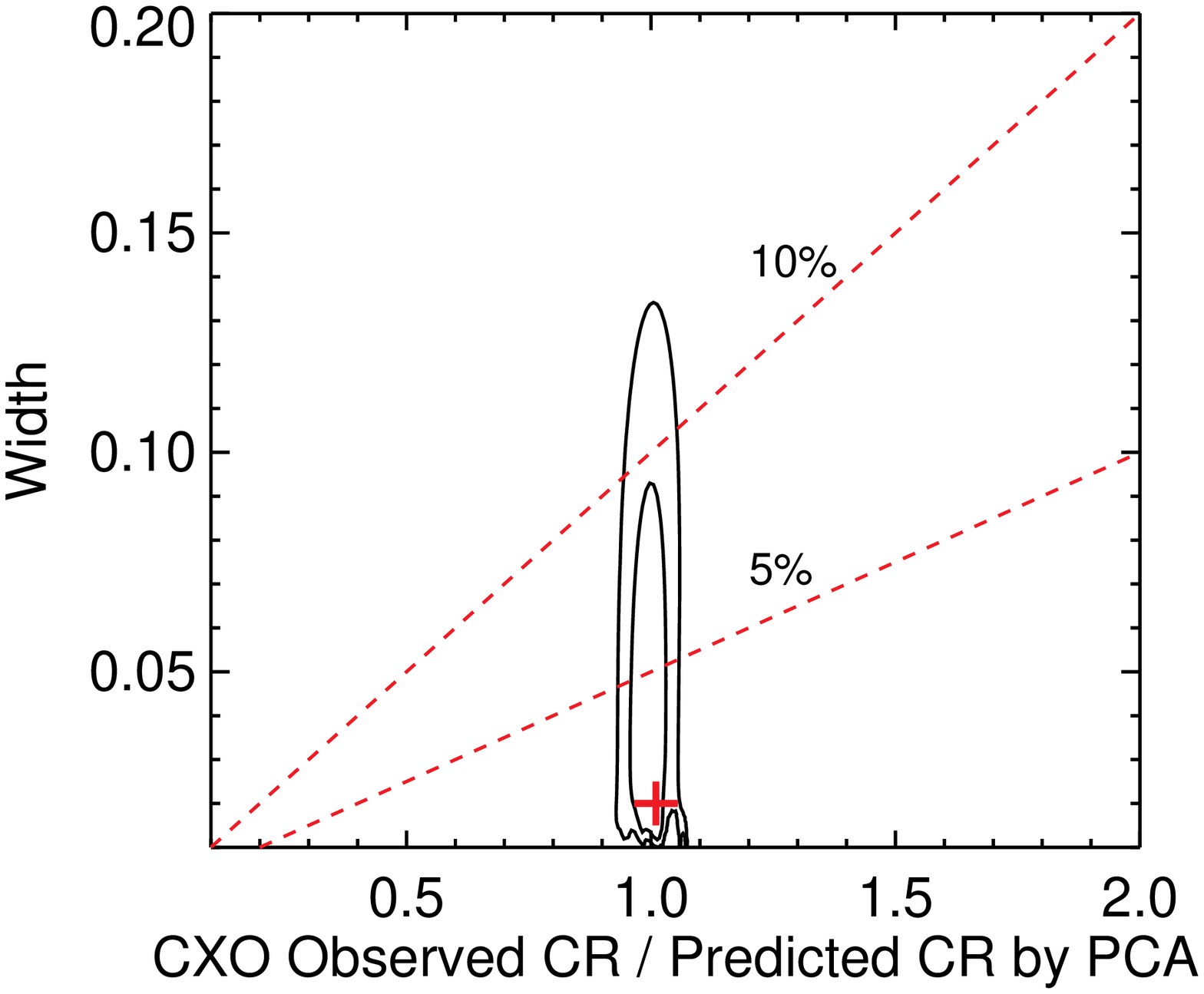}
\includegraphics[scale=0.35]{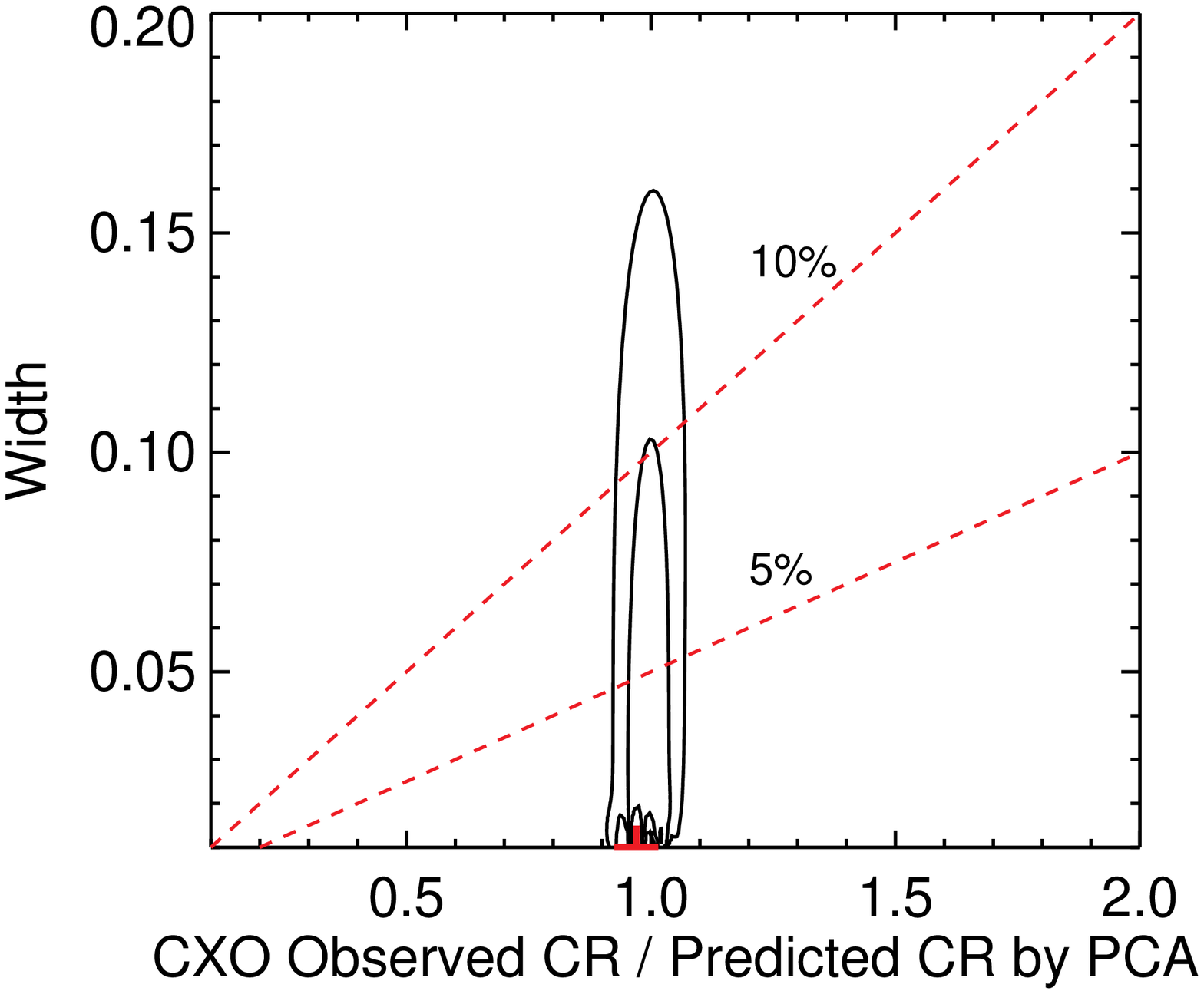}
\caption{68\%  and 95\% confidence  contours of  the parameters  of an
  assumed  underlying Gaussian distribution  of count-rate  ratios for
  X-ray  bursts  \#22, \#23,  and  \#24  observed  with Chandra  and  RXTE
  simultaneously.  The  most likely peak of the  distribution show the
  offset  between the satellites,  while the  width of  the underlying
  distribution reflects  the systematic uncertainty  in the individual
  measurements.   The  dashed  red  lines  show  the  width  when  the
  systematic uncertainty is 5\% and 10\% of the mean. }
\label{res_ch}
\end{figure*}

\begin{figure*}
\centering
\includegraphics[scale=0.35]{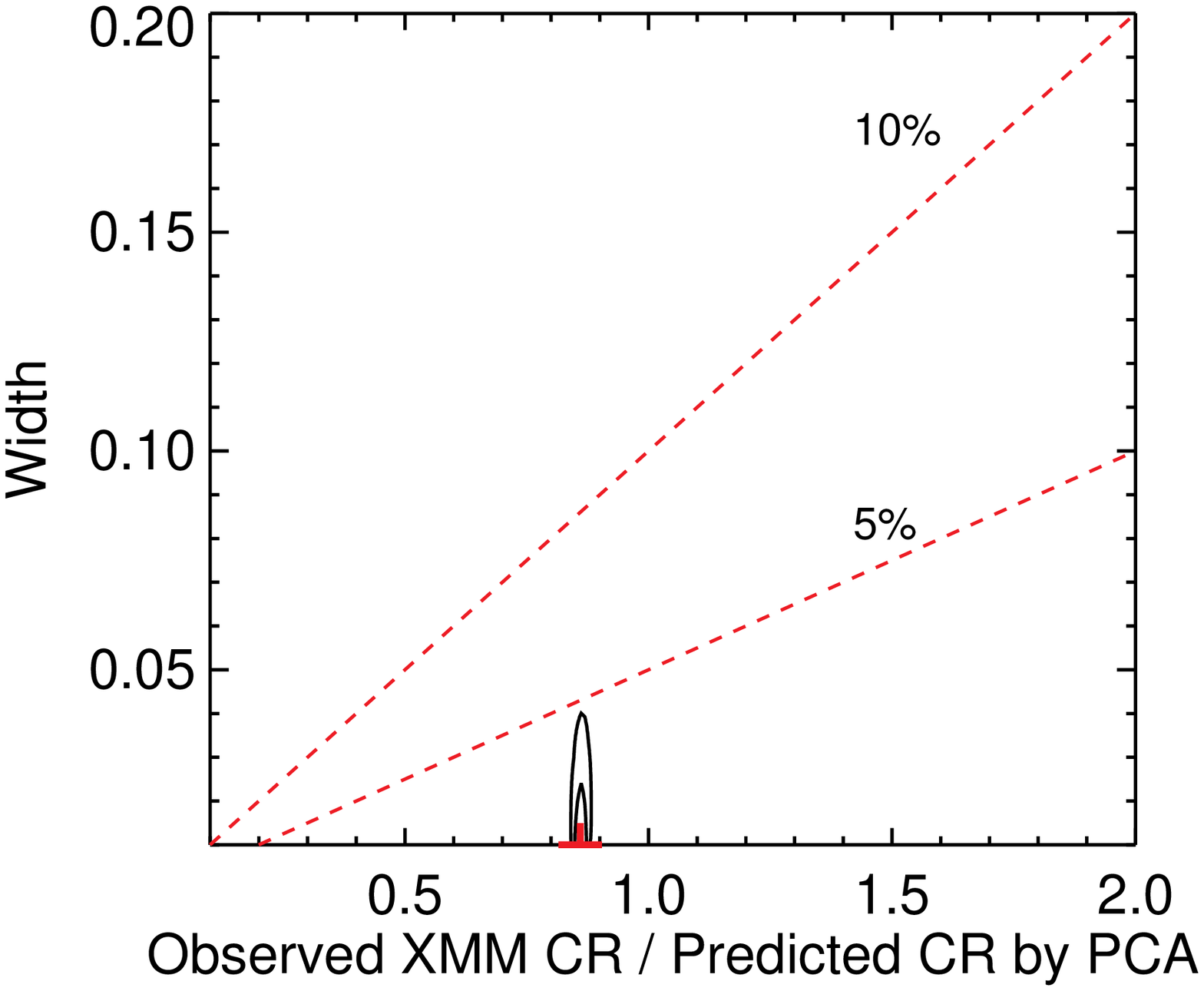}
\includegraphics[scale=0.35]{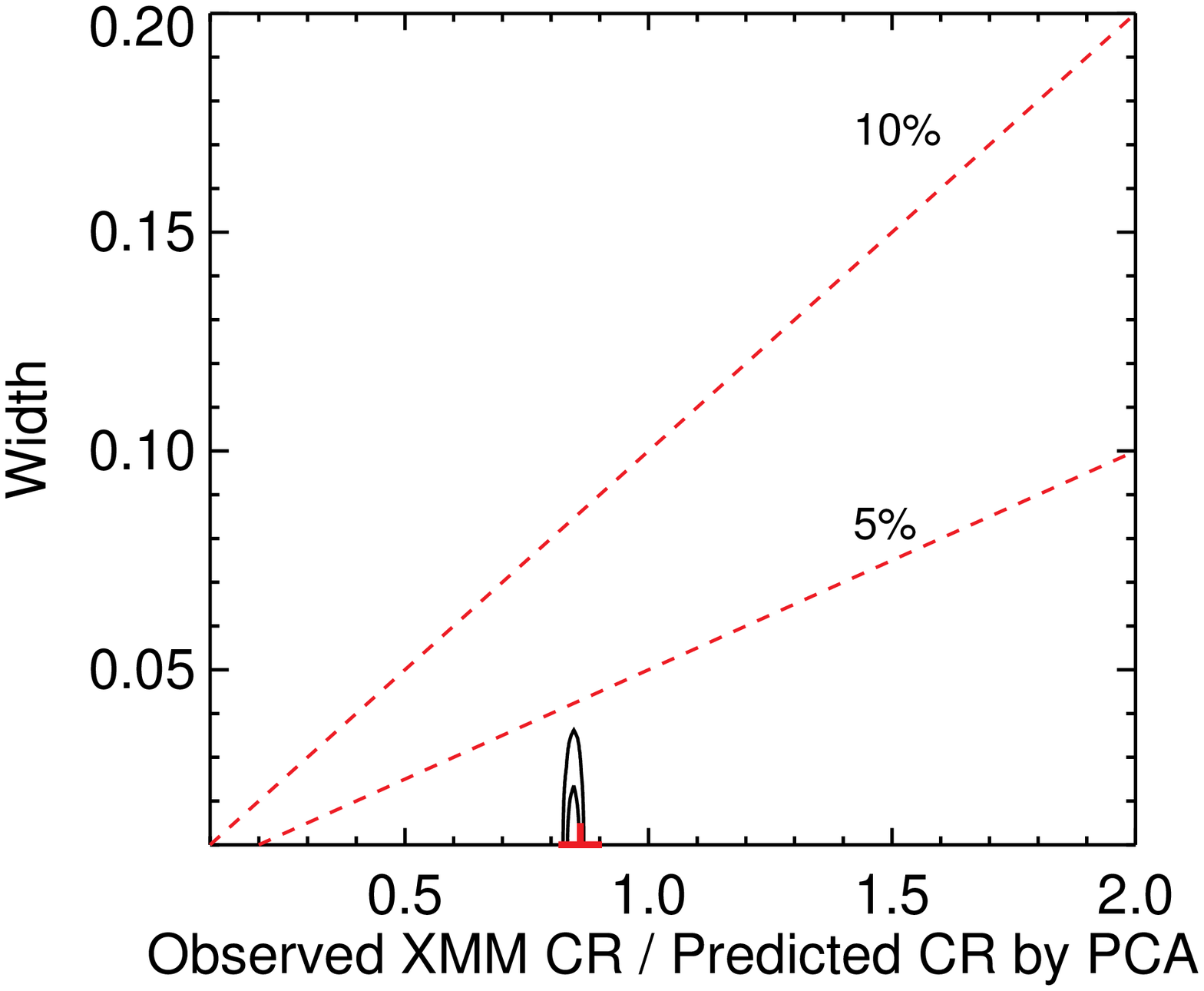}
\includegraphics[scale=0.35]{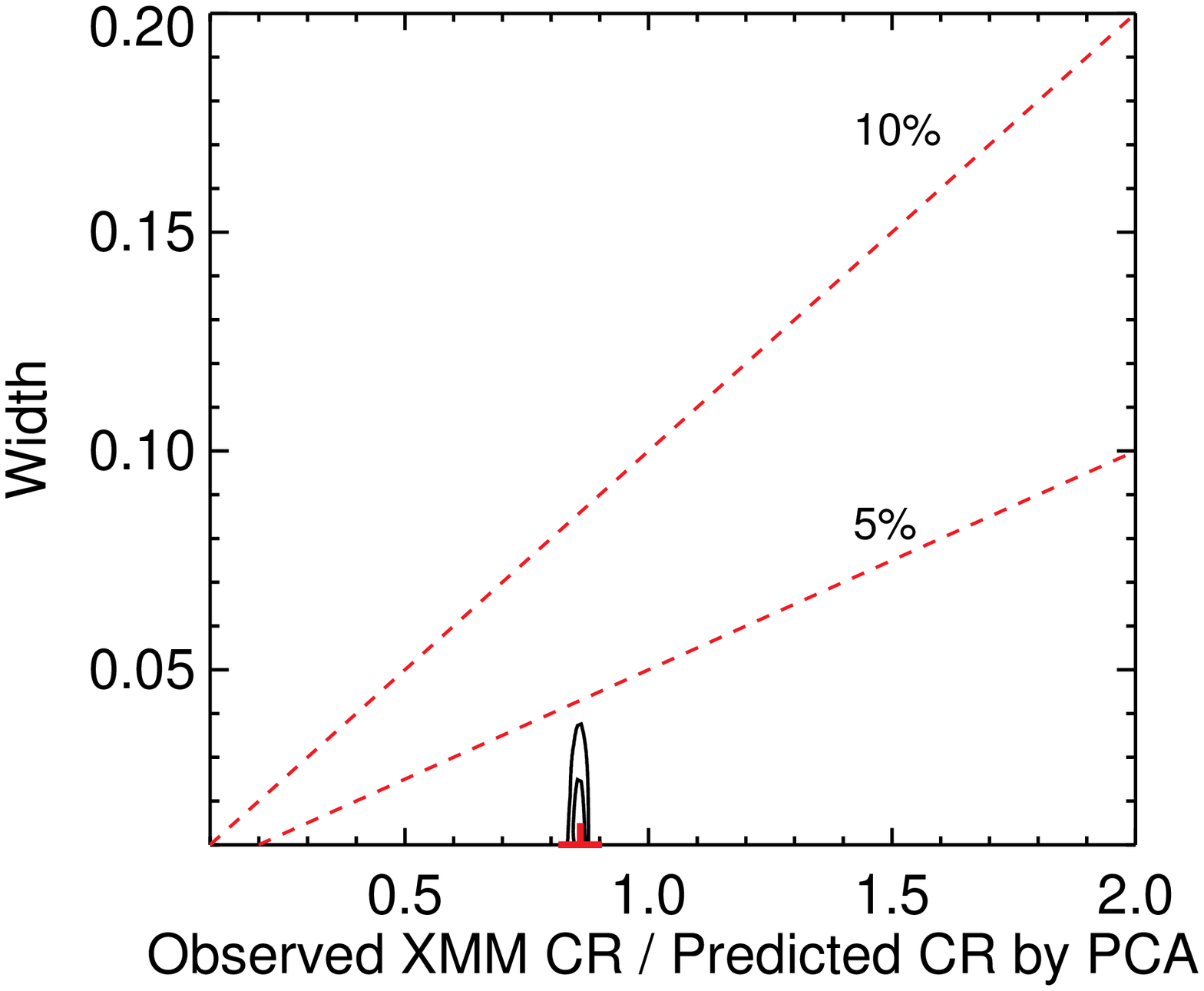}
\caption{Same as  Figure \ref{res_ch} but  for bursts \#26,  \#27, and
  \#30 observed with XMM-{\it Newton} and RXTE simultaneously.}
\label{res_xmm1}
\end{figure*}

\begin{figure*}
\centering
\includegraphics[scale=0.35]{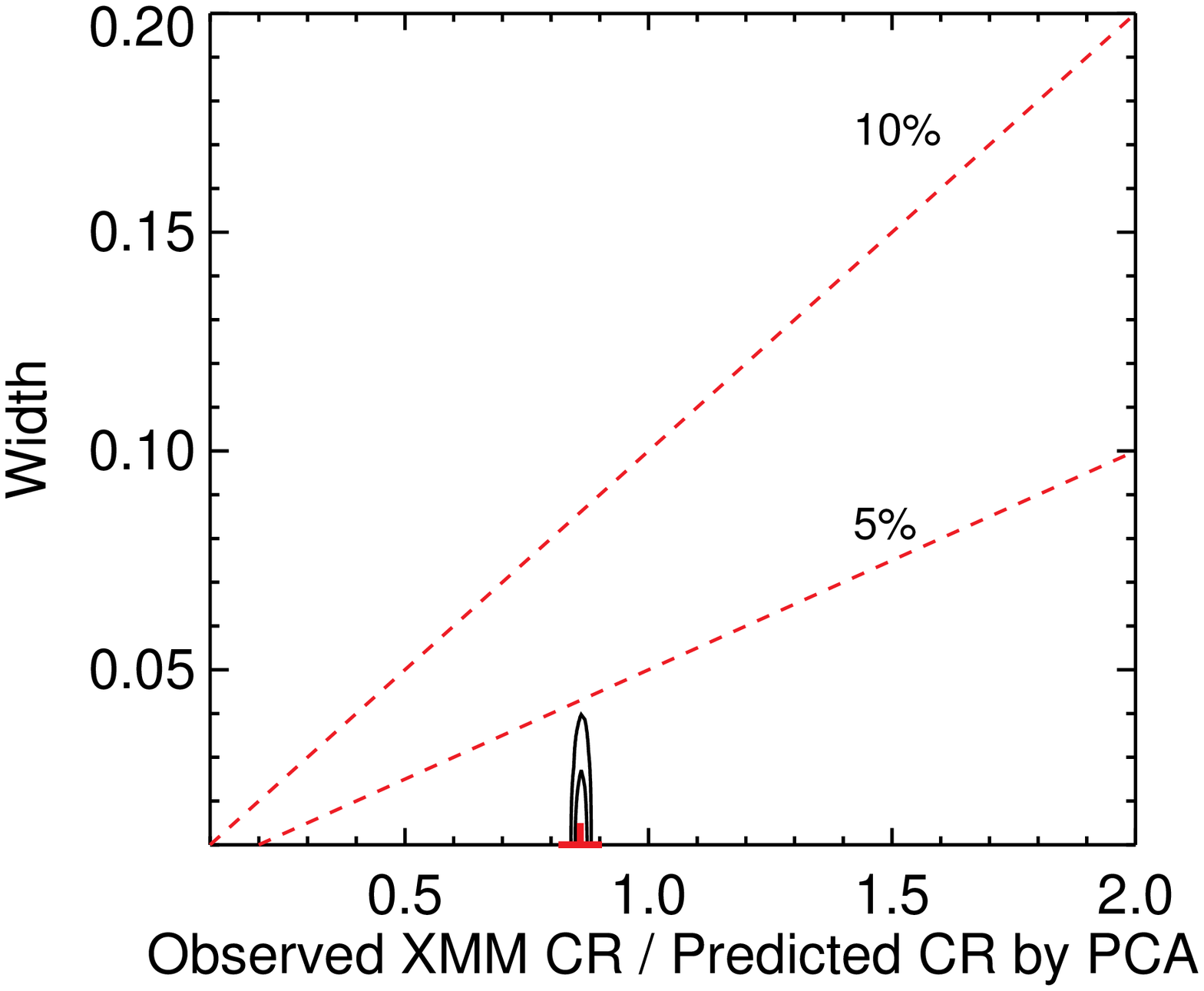}
\includegraphics[scale=0.35]{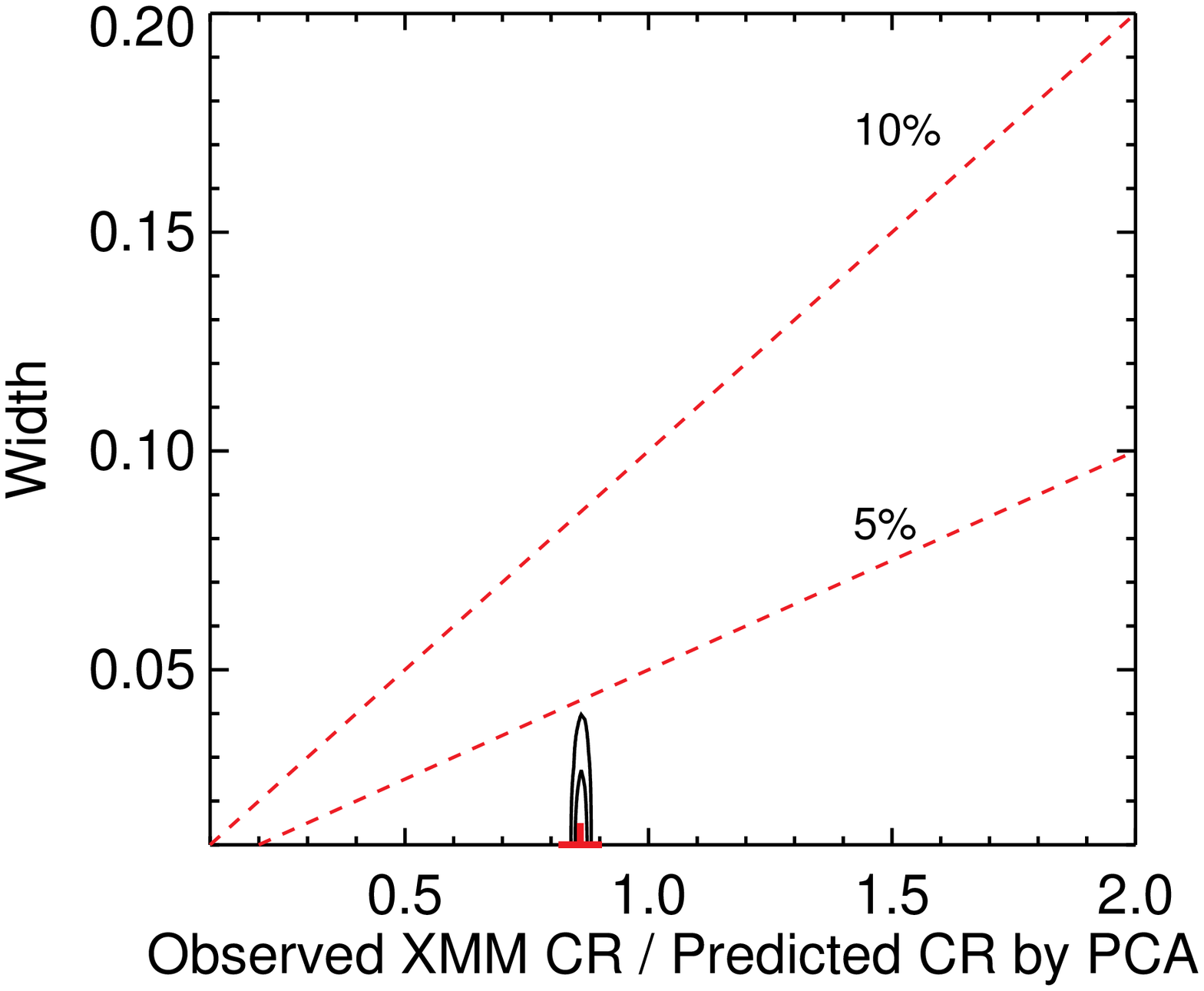}
\caption{Same as Figure \ref{res_ch} but for bursts \#31 and \#32 observed
  with XMM-{\it Newton} and RXTE simultaneously. }
\label{res_xmm2}
\end{figure*}

Using Bayes' theorem, we can write 
\begin{equation}
P(f_0,\sigma|\mbox{data})=C_0 P(\mbox{data}|f_0,\sigma) P(f_0)P(\sigma)\;.
\label{eq:bayes}
\end{equation}
Here $C_0$ is a normalization constant, and $P(f_0)$ and $P(\sigma)$
are the priors over the centroid and dispersion of the underlying
Gaussian distribution. We take these prior distributions to be flat in
the range of parameters shown in Figures~\ref{res_ch}, \ref{res_xmm1},
and \ref{res_xmm2}, which is large enough to incorporate any range of
interest.  The quantity $P(\mbox{data}|f_0,\sigma)$ measures the
likelihood that we will make a particular set of measurements given
the values of the parameters of the underlying Gaussian
distribution. Because these measurements are uncorrelated, we write
\begin{equation}
P(\mbox{data}|f_0,\sigma)=\prod_i \int df P_i(f|f_{0,i},\sigma_i)
P(f;f_0,\sigma)\;.
\label{eq:data}
\end{equation}
Inserting  equation~(\ref{eq:data}) into  equation~(\ref{eq:bayes}) we
obtain the posterior likelihood
\begin{equation}
P(f_0,\sigma|\mbox{data})=C_0 P(f_0)P(\sigma)
\prod_i \int df P_i(f|f_{0,i},\sigma_i)
P(f;f_0,\sigma)\;,
\end{equation}
where   $C_0$  is   an   overall  normalization   constant.  We   show
two-dimensional  contours  of  these  likelihoods for  each  burst  in
Figures~\ref{res_ch}, \ref{res_xmm1},  and \ref{res_xmm2}  and display
the most likely values of the parameters in Table~\ref{res_all}.

All  figures show  that the  underlying distributions  of offsets  are
extremely narrow and that the  most likely values of their dispersions
are  consistent with  being zero.   This indicates  that there  are no
systematic  problems in  the modeling  of the  energy response  of the
detectors  and  that  the  corrections  described  earlier  adequately
account  for any  pile-up  effects.   The most  likely  values of  the
centroids are  consistent with unity  for the comparison  between RXTE
and Chandra  but as low  as 0.86 for  the comparison between  RXTE and
XMM-{\it Newton}.

Because the  offsets are stable  both throughout the cooling  tails of
individual  bursts  and from  burst  to  burst,  we combined  all  the
measurements from all  the bursts observed by each  X-ray satellite to
quantify  the overall  offset and  its uncertainty.   We computed  the
posterior  likelihood for  the offset  $f$ given  the two  dimensional
posterior  likelihood  of  the  centroid value  $f_0$  and  dispersion
$\sigma$ of the underlying Gaussian distribution as
\begin{equation}
P(f)=\int \int P(f;f_0,\sigma) P(f_0,\sigma|\mbox{data}) df_0 d\sigma\;.
\end{equation}

Using 5 bursts and 447 individual  measurements and 3 three bursts and
170 individual  measurements for comparison with  XMM-{\it Newton} and
Chandra,  respectively,  we  found  that the  ratio  of  the  observed
count-rate  with the  Chandra  and XMM-{\it  Newton}  compared to  the
predictions based on the RXTE spectral analysis in the 2.5~$-$~8.0~keV
and    2.5~$-$~10~keV   range    is   1.012$^{+0.006}_{-0.021}$    and
0.857$\pm$0.003, respectively.   In other  words, the  fluxes measured
with Chandra HETG/ACIS are in  statistical agreement with the RXTE/PCA
while  the  EPIC-pn  measures  one  that  is  14\%  less.   In  Figure
\ref{final_overplot},  we   show  the  combined  likelihoods   of  the
resulting ratios of countrates between the different instruments.

\begin{deluxetable}{lcc}
\tablecolumns{4}   
\tablecaption{Measured Calibration Offsets}
\tablewidth{0pt} 
\tablehead{
\colhead{Burst ID}  & \colhead{Offset} & Instruments}
  \startdata
22 & 1.01 & HETG-ACIS/PCA \\
23 & 1.01 & HETG-ACIS/PCA \\
24 & 0.97 & HETG-ACIS/PCA \\
26 & 0.86 & EPIC-pn/PCA \\
27 & 0.85 & EPIC-pn/PCA \\
30 & 0.86 & EPIC-pn/PCA \\
31 & 0.86 & EPIC-pn/PCA \\
32 & 0.86 & EPIC-pn/PCA  
\enddata
\label{res_all}
\end{deluxetable}

\begin{figure*} \centering
\includegraphics[scale=0.35]{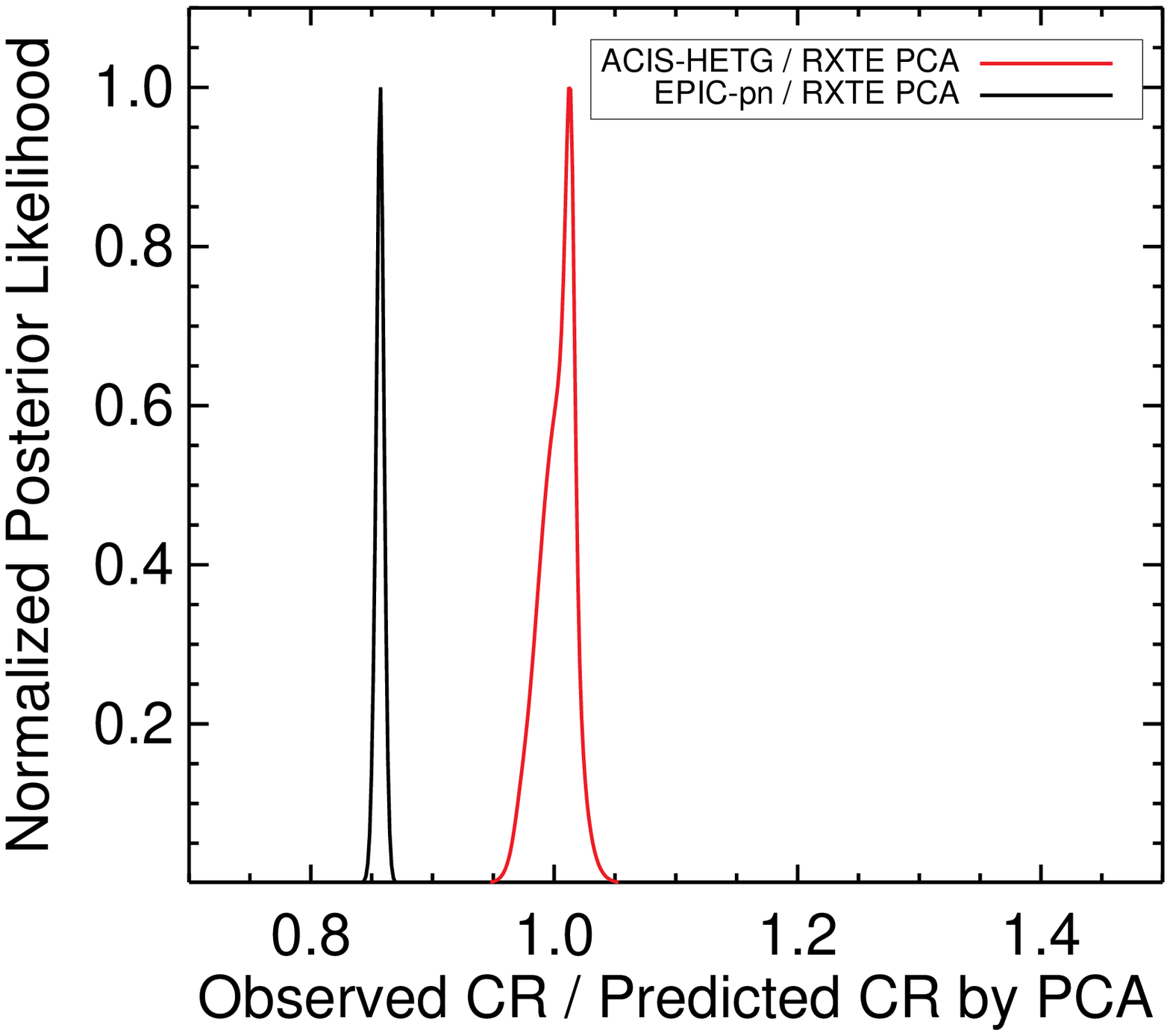}
\caption{Normalized posterior likelihoods of  the offsets derived from
  all ratio measurements for  XMM-{\it Newton} EPIC-pn/RXTE PCA (black
  solid     lines)     and     Chandra/RXTE     PCA     (red     solid
  line).} \label{final_overplot} \end{figure*}

\section{The Effect of Time-Dependent Calibration} 
\label{variable_source}
In the previous section, we addressed the cross calibration
  between RXTE and {\it Chandra} as well as between RXTE and XMM-{\it
    Newton}, using observations of the burster \src. We obtained the
  most likely fractional offsets between the fluxes measured using the
  various instruments on simultaneous observations and showed that
  they are consistent with being constant throughout individual bursts
  as well as across different bursts.  However, the offsets we
  inferred correspond to a single point in time for each comparison.
 
If the efficiencies of the detectors were constant in time, the results
of this comparison would be applicable to all the bursts observed
thoughout the durations of the missions. However, the responses of the
detectors evolve with time for a number of reasons (e.g., gas leakage,
charge transfer inefficiency). This evolution is typically accounted for
using smooth, time-dependent models of the response parameters of the
detectors. These parameters are determined by fitting the models to
regular observations of the cosmic sources with known brightness and
spectral shape or using internal calibrations sources (e.g. EPIC
cameras). As an example, RXTE/PCA had been calibrated against the Crab
nebula (Shaposnikov et al.\ 2013) throughout the mission. 

The smooth models of the evolution of the detector response clearly
cannot reliably capture the true evolution, which is expected to have
a stochastic component. This will, therefore, introduce an uncertainty
in the measurements that will change in time. {\sout This complication is
exacerbated if the calibration source shows intrinsic variability, as
is the case with the Crab Nebula, which showed variability at a level
of $\approx$~7\% (Wilson-Hodge et al.\ 2011).} It is, therefore,
possible that were we to repeat our cross calibration procedure
between RXTE and Chandra at different epochs, we would obtain
different offsets. Even though randomly catching an epoch of identical
flux measurements between RXTE and Chandra might seem unlikely, we
cannot base our conclusions on such a posterior likelihood. For this
reason, we now explore the possibility that there is a time-dependent
systematic uncertainty in the calibration of the instruments and how
this uncertainty would affect the scatter in the measured burst
fluxes.

If there were single-epoch measurements of the fluxes during the
cooling tails of bursts, this potential systematic error would have to
be added to the error budget of the measurements. However, for many of
the sources we used for neutron-star radius measurements in Papers I
and II, the burst flux measurements were carried over many epochs
spanning the lifetime of RXTE. In this case, as we will show below,
this potential calibration uncertainty should not be treated as a
separate source of systematic uncertainty as it is already inherently
part of the measured scatter of observed fluxes.

To demonstrate this point, we compute the mean and the variance of the
(burst) flux measured by an instrument (e.g., RXTE/PCA). Let
the source flux for each burst be $S_{i, \rm true}$, with a mean of
$<S_{i,\rm true}> = S_{0,\rm true}$ and an intrinsic dispersion
$\sigma_{\rm true} \ne 0$. Let the instrument have an inferred overall
efficiency factor $R$ (e.g., as reported by a calibration team). If
this efficiency factor has an unknown systematic bias, which may also
be variable in time, then the observed countrate flux will be
\begin{equation}
F_i = (R+B+ V_i) S_i, 
\end{equation}
where we have denoted the mean bias in the calibration over the time
span of the observations by $B$ and the potential time-variable part
by $V_i$. By definition, $\langle V_i\rangle = 0$.

When an observation is carried out and standard calibration procedures
are applied, the observed source flux will be inferred to be equal to
\begin{equation}
  S_{i, \rm obs} = \frac{F_i}{R},
\end{equation}
which will be related to the true source flux by
\begin{equation}
  S_{i, \rm obs} = \frac{R+B+V_i}{R} S_{i, \rm true}. 
\end{equation}
Clearly, the observations reproduce the true source flux if
there is no bias in the calibration. 

The mean inferred flux of the source is equal to
\begin{equation}
  S_0 = \langle S_{i, \rm obs} \rangle = \left(1+\frac{B}{R}\right)
\langle S_{i, \rm true} \rangle,   
\end{equation}
where we have used the fact that $\langle V_i S_{i, \rm true} \rangle
= 0$ because the time evolution of the detector response and the
variability of the source are uncorrelated. The variance in the
inferred burst flux is
\begin{eqnarray}
\sigma_{\rm obs}^2 &=& \langle (S_{i,\rm obs} - S_0)^2 \rangle \nonumber \\
           &=& \langle S_{i,\rm obs}^2 + S_0^2 - 2 S_{i,\rm obs} S_0 \rangle \nonumber \\  
	   &=& \langle S_{i,\rm obs}^2 \rangle - S_0^2
\end{eqnarray}
and the mean of the square of the inferred fluxes is given by
\begin{eqnarray}
  \langle S_{i,\rm obs}^2 \rangle &=& \left\langle
  \left[\left(1+\frac{B}{R}+\frac{V_i}{R}\right) S_{i, \rm true}\right]^2
  \right\rangle \nonumber \\
  &=& \left(1+\frac{B}{R}\right)^2 \langle
  S_{i,\rm true}^2\rangle + \left\langle \left(\frac{V_i}{R}\right)^2
  S_{i,\rm true}^2 \right\rangle + 2\left(1+\frac{B}{R}\right) \left\langle
  \frac{V_i}{R} S_{i,\rm true}^2 \right\rangle.
\end{eqnarray}
Since $\langle V_i S_{i,\rm true}^2 \rangle = \langle V_i\rangle
\langle S_{i,\rm true}^2\rangle=0$, we find that
\begin{eqnarray}
  \sigma_{\rm obs}^2&=&
  \left(1+\frac{B}{R}\right)^2 \langle
  S_{i,\rm true}^2\rangle + \left\langle \left(\frac{V_i}{R}\right)^2
  S_{i,\rm true}^2 \right\rangle
  -\left(1+\frac{B}{R}\right)^2\langle S_{i, \rm true} \rangle^2\nonumber\\
  &=&
  \left(1+\frac{B}{R}\right)^2
  \left[\langle S_{i,\rm true}^2\rangle  -\langle S_{i, \rm true} \rangle^2\right]
    + \frac{1}{R^2}\left\langle V_i^2S_{i,\rm true}^2 \right\rangle \;.
\end{eqnarray}
The term in the square bracket is equal to the true variance of the
source flux $[\langle S_{i,\rm true}^2 \rangle - \langle S_{i,\rm
    true}\rangle^2] = \sigma_{\rm true}^2$. The average in the
second term is $\langle V_i^2 S_{i,\rm true}^2 \rangle = \sigma_V^2
\langle S_{i,\rm true}^2\rangle$, where we have used the definition of the
variance of the time-variable bias
\begin{equation}
  \sigma_V^2\equiv \langle V_i^2\rangle -\langle V_i\rangle^2=\langle V_i^2\rangle\;.
\end{equation}
As a result,
\begin{equation}
  \sigma_{\rm obs}^2 = \sigma_{\rm true}^2  \left(1+\frac{B}{R} \right)^2
  + \frac{\sigma_V^2}{R^2} \langle S_{i,\rm true}^2\rangle
\end{equation}
or in terms of fractional errors,
\begin{eqnarray}
  \left( \frac {\sigma_{\rm obs}}{S_0} \right)^2 &=& \left( \frac {\sigma_{\rm true}}
       {\langle S_{i,\rm true}\rangle} \right)^2 +\frac{\sigma_V^2}{R^2+B^2}
\frac{\langle S_{i,\rm true}^2\rangle}{\langle S_{i,\rm true}\rangle^2}\nonumber\\
&=&\left( \frac {\sigma_{\rm true}}{\langle S_{i,\rm true}\rangle}\right)^2 +\frac{\sigma_V^2}{(R+B)^2}\left[1+
  \left(\frac{\sigma_{\rm true}}{\langle S_{i,\rm true}\rangle}\right)^2\right]\;.
\end{eqnarray}
Keeping only terms up to second order in the dispersions, the final 
result is 
\begin{equation}
\left( \frac {\sigma_{\rm obs}}{S_0} \right)^2 = 
\left( \frac {\sigma_{\rm true}}{\langle S_{i, \rm true}\rangle} \right)^2 +\frac{\sigma_V^2}{(R+B)^2}\;.
\end{equation}
This expression shows that the measured fractional variance in the
burst fluxes is equal to the quadrature sum of the intrinsic
fractional variance and the fractional variance of the calibration
bias.  In other words, the observed variance in the burst fluxes is an
upper limit to the intrinsic scatter. If the statistical properties of
the calibration uncertainties could be measured independently, this
contribution could be subtracted from the dispersion of measured
fluxes.

\section{Conclusions}

In this paper,  we quantified the systematic  uncertainties that arise
from the  flux calibration  of the X-ray  instruments used  in neutron
star mass  and radius measurements. We  used simultaneous observations
of the burster \src\ with RXTE/PCA  and Chandra HETG/ACIS-S as well as
with RXTE/PCA  and XMM-{\it  Newton}/EPIC-pn. We performed  a Bayesian
statistical  analysis to  measure  the  level of  offset  in the  flux
calibration between these instruments.

In the case of the comparison between Chandra and RXTE, we found no
evidence for a significant offset between the measured fluxes and concluded 
that the most likely value of the flux ratio is
$1.012^{+0.006}_{-0.021}$. As we describe in the appendix, the effect
of pileup is unlikely to change the value of this ratio in a statistically 
significant way. In contrast, we found a 14.0$\pm$0.3\%
difference between the fluxes inferred using XMM-{\it Newton}/EPIC-pn
and RXTE/PCA. This discrepancy is persistent in all five analyzed
bursts and is independent of the source spectrum or countrate.

The flux discrepancy between XMM-{\it Newton}/EPIC-pn and other
detectors has been noted in earlier studies of different
sources. Nevalainen et al.\ (2010) reported a difference between flux
measurements with the EPIC-pn, Chandra ACIS (zeroth-order data), and
EPIC MOS detectors at a level of 5$-$10\%.  According to their
results, ACIS flux measurements are significantly higher than EPIC-pn,
by 11.0$\pm$0.5\%.  Similarly, the difference in the flux measurements
between the EPIC-pn and MOS detectors has been noticed earlier (see
e.g., Mateos et al.\ 2009 and XMM-Newton calibration documents
XMM-Newton-SOC-CAL-TN-0018\footnote[3]{http://xmm2.esac.esa.int/docs/documents/CAL-TN-0018.pdf}
and
XMM-Newton-SOC-CAL-TN-0052\footnote[4]{http://xmm2.esac.esa.int/docs/documents/CAL-TN-0052.ps.gz});
MOS yields fluxes that are approximately 7\% higher than EPIC-pn (see
also Read et al. 2014). Most recently, using 64 clusters of galaxies,
Schellenberger et al.  (2014) showed that the effective area
cross-calibration uncertainty between ACIS and EPIC-pn is energy
dependent and in the 2$-$7~keV range stabilizes at a 15\% level a
similar result is also presented by Guainazzi et al. (2015).

Given that the calibration errors of  the detectors may be variable in
time, it  is possible  that the excellent  agreement between  RXTE and
Chandra is a coincidence.  We showed, however, that even in this case,
the spread  of the measured burst  fluxes reported in Papers  I and II
already  accounts  for  this potential  systematic  uncertainty.   We,
therefore, conclude that, while an overall bias may still exist in the
measurements, potential time-dependent  flux calibration uncertainties
do  not  increase the  error  budget  of  the radius  measurements  we
reported in these earlier papers.

One final consideration is whether the difference in the measured flux
between XMM-{\it Newton} EPIC-pn and the other two instruments also
affects the neutron star radii determined through the observations of
quiescent low-mass X-ray binaries.  The flux levels in question are
very different in the observations of these dim sources: they are
approximately six orders of magnitude below the peak burst fluxes and
the Crab nebula, and about three orders of magnitude below other
typical calibration sources (e.g., clusters of galaxies).  Given the
signal-to-noise ratio of the spectra that is achieved with even the
longest observations of quiescent low-mass X-ray binaries, it is not
possible to discern any measured flux differences between instruments,
even if this difference persisted at the 14\% level. This supports the
lack of evidence for flux differences presented in the earlier studies
(Catuneanu et al.  2013; Heinke et al.  2014; Guillot et al.  2013),
which compared neutron star radii derived from the quiescent sources
using XMM-{\it Newton} and Chandra data.

\appendix
\section{Correcting for pileup in Chandra HETG/ACIS-S}
To account for pileup in the Chandra HETG/ACIS-S spectra, we used a
correction that was developed independently by one of the co-authors,
Herman Marshall, for mildly piled-up calibration data.  The approach
relies on the measurement of counts per pixel around the Ir-M and Si-K
edges, at which the spectral response changes rapidly.  Denoting the
observed counts per detector pixel per frame in wavelength bin $i$ as
$R_i$, we define the degradation of the effective area as

\begin{equation}
  A_i^{\prime} = A_i \exp(-a R_i),
  \label{eqa1}
\end{equation}
 where the exponential model was chosen due to the Poisson nature
  of the probability of multiple events for a given count rate.

  The value of the parameter $a$ and the uncertainty associated with
  its determination was found using an observation of Mk~421
  (observation ID 4148) performed on 26$-$27 October 2002 with the
  LETGS/ACIS-S for over 91~ks.  The data were downloaded from {\it
    TGCat}\footnote[2]{http://tgcat.mit.edu} (Huenemoerder et al.
  2011).  The countrate spectra were rebinned by a factor of two so
  that each spectral bin corresponds closely to one ACIS pixel.  The
  source was extremely bright so that there were of order 5000 counts
  per spectral bin in both the +1 and -1 orders near 2~keV.  The count
  spectrum was divided by the ARF obtained from TGCat to generate
  spectra in photons/cm$^{-2}$/s/\AA\ for both the $+1$ and $-1$
  orders.  When the two orders are compared (see the top panel of
  Figure~\ref{final_overplot}) without being corrected for pileup,
  there are clear edges at Ir-M (6.0~\AA) and Si-K (6.74~\AA) in the
  effective-area-corrected spectra.  These features arise from
  differential pileup, where $R_i$ may increase significantly over a
  few spectral resolution elements at an edge in the effective area of
  the instrument, causing increased pileup and a larger fractional
  loss of events with a corresponding decrease in apparent photon
  flux.

  We devised a model-independent method for estimating the parameter
  $a$ in equation~(\ref{eqa1}), which depends on the very smooth and
  slowly varying photon spectrum of Mk~421.  Using the ratio of fluxes
  in the 5.5-5.9~\AA\ range to those in the 6.1-6.5 \AA\ range
  (straddling the Ir-M edge), we compute the value of $a$ that matches
  the ratios based on the average $R$ values in the two narrow bands.
  We find $a = 4.4$ for $-1$ and $a=5.5$ for $+1$ and, therefore,
  adopt $a = 5$ and a range of acceptable values that are $+/- 0.5$
  away from this value.  Figure~\ref{final_overplot} shows the
  sensitivity of the spectral inversion to $a$; for $a=4$, the spectra
  still show edges, while, for $a=7$, the data are over-corrected to
  form an excess over a smooth spectral model.  (A global spectral fit
  to a double power law indicates that the photon spectrum should drop
  monotonically by only 1\% per \AA\ through this wavelength range, so
  the assumption that the fluxes are equal should be good to 1\%.)

Using equation~(\ref{eqa1}), which is verified to $R_i = 0.05$
cts/frame/pixel, a $\pm 10$\% uncertainty in $a$ results in at most a
$\pm 2.5$\% uncertainty in $A_i^{\prime}/A_i$, which is smaller than
the expected calibration uncertainty in $A_i$.  This method required
the HRMA effective area correction implemented in 2005, which
corrected the Ir-M edge caused by a thin contaminant layer on the
HRMA. Note that, the correction given in equation~(\ref{eqa1}) is
valid for low to modest pileup, i.e., $R_i < 0.2$~cts/frame/pixel and
the data used to compute the coefficient $a$ ranged up to about 0.10
cts/s/pixel. This yields a maximum correction of about 50\%.

In our comparison between RXTE and Chandra, we took into account the
pileup losses in the predicted countrates rather than correcting the
observed rates for pileup.  If we denote by $y_i$ the number of
photons per detector pixel per frame that should have been recorded in
wavelength bin $i$ in the absence of pileup, we can then invert the
above expression to obtain

\begin{equation}
R_i = y_i \left[ 0.63 \exp\left(-6 y_i\right) + 0.37\right], 
\end{equation}

This approximation is appropriate for $y_i < 0.065$ cts/frame/pix.
The burst spectra reached 0.08 cts/frame/pix, which is a small
extrapolation from the domain for which the correction was derived.

\begin{figure*} \centering
\includegraphics[scale=0.35]{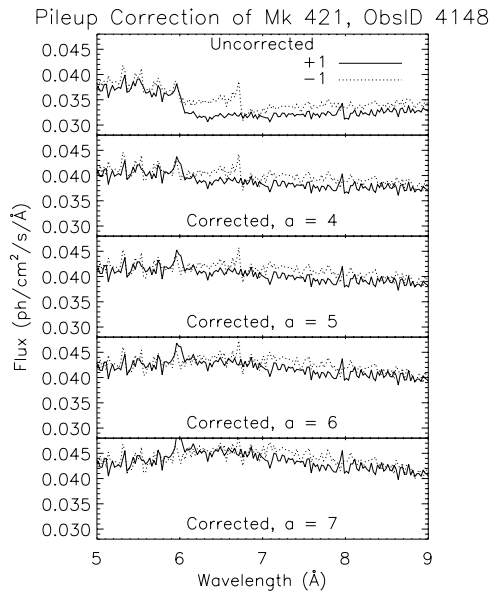}
\caption{The 5$-$9~\AA\ region of the unfolded photon spectra of
  Mk~421 (ObsID 4148), taken with the {\em Chandra} LETG/ACIS-S.  The
  top panel shows the original data, uncorrected for pileup, as
  obtained from the tgcat archive of {\em Chandra} grating spectra.
  The remaining panels show how the spectra appear after adjusting the
  effective area according to equation~(\ref{eqa1}), for several
  values of the parameter $a$.  It is clear that the uncorrected
  spectra show spectral jumps at the instrumental Ir-M (6.0 \AA) and
  Si-K (6.74 \AA) edges.  The spectra are most consistent with a
  simple smooth power law for $a = 5$.} 
\label{final_overplot} 
\end{figure*}

\acknowledgements{We thank the members of both the Chandra and
  XMM-Newton help desks for their help and suggestions throughout this
  study. We also thank Tod Strohmayer for his encouragement to pursue
  a study of flux calibration uncertainties in neutron star radius
  measurements. We thank Keith Jahoda for illuminating discussions on
  the calibration of the RXTE/PCA instrument. TG was supported by
  Scientific Research Project Coordination Unit of Istanbul
  University. Project numbers: 49429, 57321 and 48934. DP and FO gratefully
  acknowledge support from NASA ADAP grant NNX12AE10G and NSF grant
  AST-1108753 for this work.  This research has made use of data
  obtained from the High Energy Astrophysics Science Archive Research
  Center (HEASARC), provided by NASA's Goddard Space Flight Center.}

\end{document}